\documentclass{iopart} 

\expandafter\let\csname equation*\endcsname\relax
\expandafter\let\csname endequation*\endcsname\relax
\usepackage{amsmath,amssymb,amsfonts}
\usepackage{comment}
\usepackage{cite}
\usepackage{algorithmic}
\usepackage{graphicx}
\usepackage{textcomp}
\usepackage{xcolor}
\usepackage{xspace}
\usepackage{url}
\usepackage{hyperref}
\usepackage{breakurl}
\usepackage{soul}
\usepackage{multirow}

\newcommand{\Feqgpu}{\ensuremath{F^{\text{eq}}_{\text{GPU}}}\xspace}
\newcommand{\unit}[1]{\ensuremath{\text{\,#1}}\xspace}
\usepackage[final]{fixme}
\fxsetup{layout=footnote, marginclue}

\begin{document}
\begin{flushright}
FERMILAB-PUB-20-338-E-SCD
\end{flushright}

\title[GPU coprocessors as a service for deep learning inference in high energy physics]{GPU coprocessors as a service for deep learning inference in high energy physics}

\author{Jeffrey Krupa$^{1}$, Kelvin Lin$^{2}$, Maria Acosta Flechas$^{3}$, Jack Dinsmore$^1$, Javier Duarte$^{4}$, Philip Harris$^{1}$, Scott Hauck$^{2}$, Burt Holzman$^{3}$, Shih-Chieh Hsu$^{2}$, Thomas Klijnsma$^{3}$, Mia Liu$^{3}$, Kevin Pedro$^{3}$, Dylan Rankin$^{1}$, Natchanon Suaysom$^{2}$, Matt Trahms$^{2}$, Nhan Tran$^{3,5}$ }
\address{$^{1}$ Massachusetts Institute of Technology, Cambridge, MA 02139}
\address{$^2$ University of Washington, Seattle, WA, 98195}
\address{$^3$ Fermi National Accelerator Laboratory, Batavia, IL 60510}
\address{$^4$ University of California San Diego, La Jolla, CA 92093}
\address{$^5$ Northwestern University, Evanston, IL 60208}


\begin{abstract}
In the next decade, the demands for computing in large scientific experiments are expected to grow tremendously.
During the same time period, CPU performance increases will be limited. 
At the CERN Large Hadron Collider (LHC), these two issues will confront one another as the collider is upgraded for high luminosity running. 
Alternative processors such as graphics processing units (GPUs) can resolve this confrontation provided that algorithms can be sufficiently accelerated. 
In many cases, algorithmic speedups are found to be largest through the adoption of deep learning algorithms. 
We present a comprehensive exploration of the use of GPU-based hardware acceleration for deep learning inference within the data reconstruction workflow of high energy physics. 
We present several realistic examples and discuss a strategy for the seamless integration of coprocessors so that the LHC can maintain, if not exceed, its current performance throughout its running.
\end{abstract}

\newcommand{\MLST}{\emph{Mach. Learn.: Sci. Technol.}}   
\submitto{\MLST}{\href{https://doi.org/10.1088/2632-2153/abec21}{doi:10.1088/2632-2153/abec21}}


\section{Introduction}
The detectors at the CERN Large Hadron Collider (LHC)~\cite{Evans:2008zzb} have enormous data rates, with a current aggregate of 100\unit{Tb/s} and plans to exceed over 1\unit{Pb/s}.
The challenge of processing this data continues to be one of the most critical elements in the execution of the LHC physics program~\cite{Boyd:2020qox,Gerber:2019jlx,Dainese:2019rgk,Alimena:2019zri,Abercrombie:2015wmb}. 
A three-tiered approach is utilized to process LHC data, where at each tier, the data rate is reduced by roughly two orders of magnitude, resulting in a manageable final data rate of 10\unit{Gb/s}.
Due to the high initial rate and restrictions coming from the high radiation collision environment, the first tier of computing consists of specialized hardware that utilizes field-programmable gate arrays (\mbox{FPGAs}) and application-specific integrated circuits (ASICs). 
The second tier, the high-level trigger (HLT), consists of a CPU-based computing cluster on-site at the LHC.
The first two tiers are described as ``online'' computing, because they occur in real time, as each LHC collision is measured by the detector.
The third tier, performing complete event processing, consists of a globally distributed CPU-based computing grid.
This third tier is described as ``offline'' computing, because it occurs after the initial collision data has already been written to disk.
Both the online and offline computing tiers run a similar set of algorithms, but the HLT employs certain approximations in order to satisfy the online latency budget.

\begin{figure}[htbp]
\centering
\includegraphics[keepaspectratio, width=0.85\columnwidth]{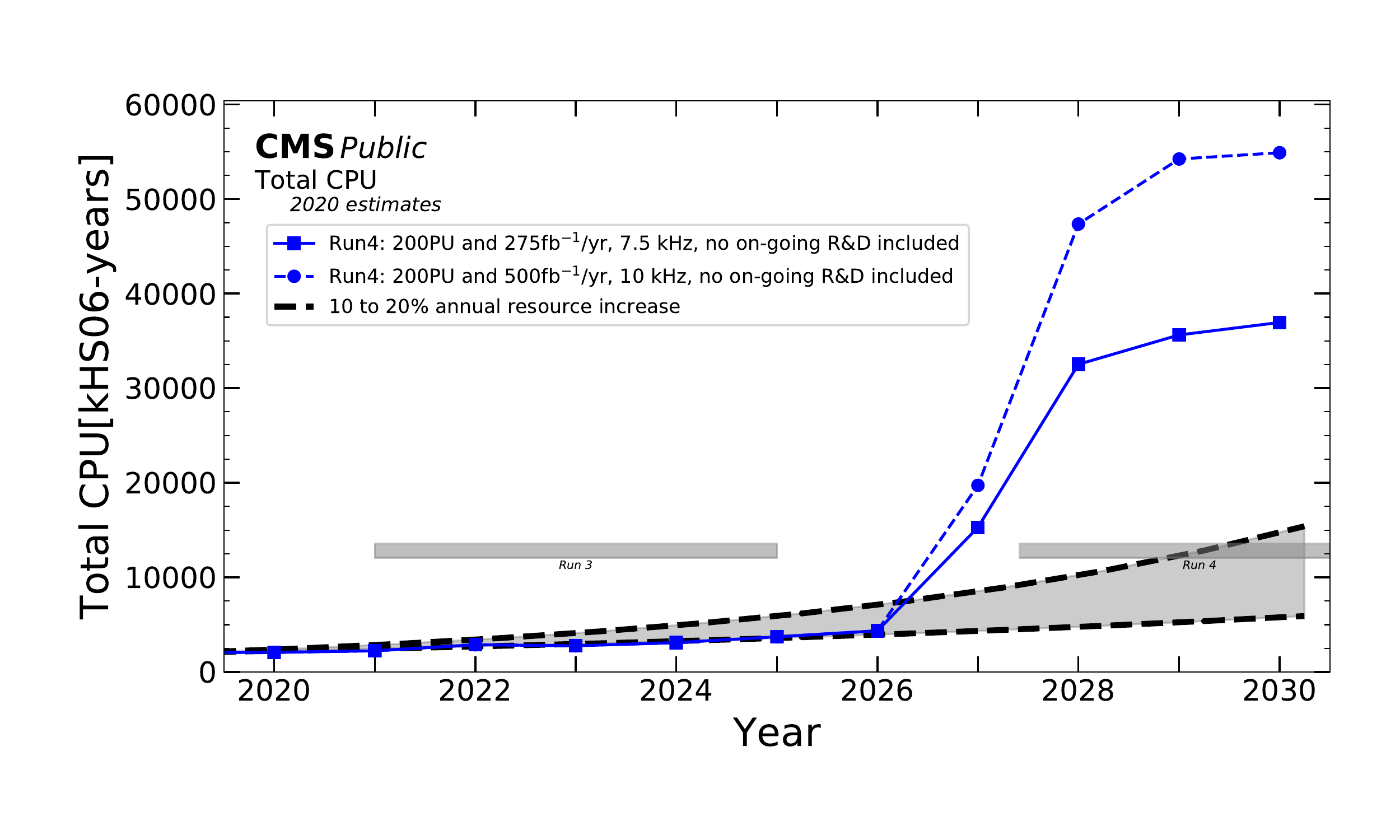}
\includegraphics[keepaspectratio, width=0.85\columnwidth]{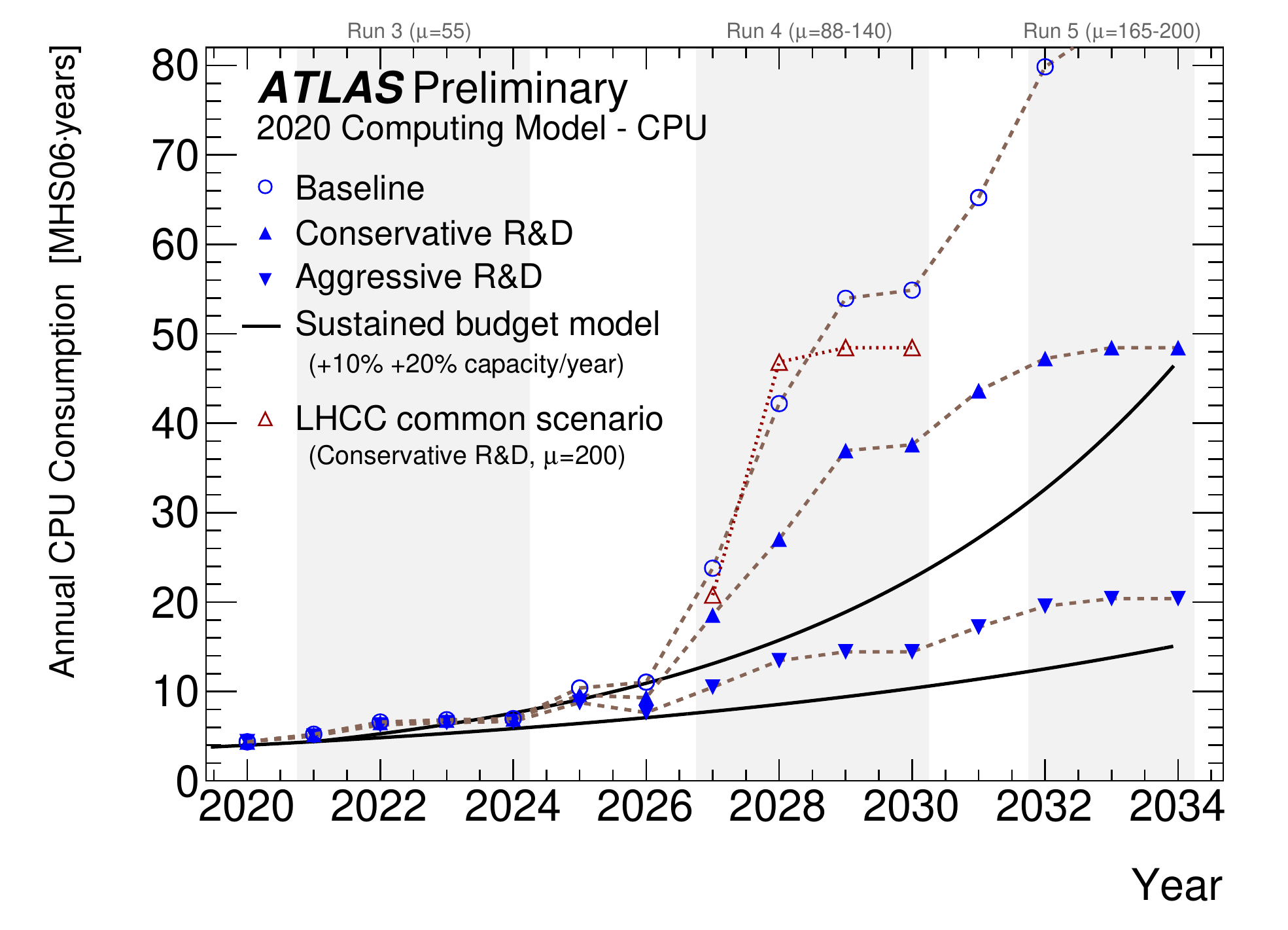}
\caption{Computing needs projections for the CMS (upper) and ATLAS (lower) experiments at the LHC~\cite{Albrecht_2019,octwiki,ATLASCompute}. 
For ATLAS, the computing needs estimate is shown under the baseline computing model and under different R\&D scenarios, while for CMS, the estimate is shown without including ongoing R\&D in different collider conditions.
The black solid lines show the range of expected available CPU, assuming CPU performance increases of 10--20\% per year.
MHS06-years (kHS06-years) stands for $10^{6}$ ($10^3$) HEPSPEC06 per year, a standard CPU performance metric for high energy physics~\cite{HS06}. 
The different LHC operating runs are also indicated.
A large upgrade to the collider and all LHC detectors will occur between Run 3 and Run 4, starting in 2026.}
\label{fig:compute}
\end{figure}

The first decade of LHC running has led to an extensive set of scientific results. 
These results include the discovery of the Higgs boson~\cite{:2012gk,:2012gu,Chatrchyan:2013lba} and, more recently, strong constraints on the nature of dark matter~\cite{Sirunyan:2017jix,Sirunyan:2019vgj,Sirunyan:2019vxa}. 
To contend with these strong dark matter constraints, physicists have been forced to re-think their approach to searching for dark matter and, generically, new physics models. 
This has led to the development of light dark matter models~\cite{alex2016dark}. 
These models often predict signatures that could be produced at the LHC but would be discarded in the early tiers of data reduction. 
To enable the search for these particles, it is imperative to increase the probability that these particle signatures are not discarded by the data reduction. 
This can be done by improving the quality of LHC data reconstruction at all tiers of processing. 
Additionally, over the next decade, the LHC will progressively increase the beam intensity, resulting in more data recorded by the detectors~\cite{Dainese:2019rgk}.
As a consequence, the demands for computing will increase proportionally to sustain the current level of physics output. 
Figure~\ref{fig:compute} shows the expected computing needs over the next decade.
These expectations arise from modeling by the two general-purpose particle detectors at the LHC: the Compact Muon Solenoid (CMS) and ATLAS experiments.
To contend with the high-luminosity upgrade of the LHC (HL-LHC), a large increase is needed starting from 2026. 
These demands outpace the expected growth of CPU performance. 
As a consequence, the LHC needs a computing solution at least to sustain the current computing performance and, potentially, to exceed it.

With the end of Dennard scaling~\cite{dennard} in the late 2000s, processor technology has undergone several changes~\cite{breakdown}. 
These changes have included the adoption of multicore processors and the rise of alternative processing architectures, or coprocessors, such as graphics processing units (GPUs), FPGAs, and ASICs. 
With the rise of deep learning (DL), these alternatives have become increasingly appealing due to the inherent parallelism in both DL algorithms and in these coprocessors. 
The gains from using coprocessors can be substantial, with improvements in inference latency for large algorithms exceeding multiple orders of magnitude~\cite{Duarte:2019fta}.
Given the scale of developments related to DL, future growth in processor technology is increasingly leaning towards heterogeneous systems in which combinations of CPUs, GPUs, FPGAs, and ASICs are all deployed, with each designed to solve specific tasks. 
However, high energy physics (HEP) experiments have thus far undertaken only limited use of alternative processors within the HLT and offline computing grids, despite common use of machine learning (ML).
HEP experiments have historically relied on ML as a way to improve the overall quality of the data and to separate small signals from enormous backgrounds, such as the discovery of the Higgs boson~\cite{Guest_2018}. 
DL approaches have enhanced both the performance and flexibility of ML techniques. 
In light of this, the LHC experiments have been quick to adopt DL techniques to improve the quality of data analysis especially during Run 2 (2015--2018)~\cite{Guest_2018,Albertsson:2018maf,Bourilkov:2019yoi,Larkoski:2017jix}. 
This includes core components such as low-level detector energy reconstruction~\cite{Qasim:2019otl}, electron and photon reconstruction~\cite{Belayneh:2019vyx}, and quark and gluon identification~\cite{Komiske:2016rsd,ATL-PHYS-PUB-2017-017,Andrews:2019faz}. 
The increasing deployment of these algorithms is starting to comprise a significant portion of the overall computing budget: the full event reconstruction takes tens of seconds per event on modern CPUs~\cite{WLCGmtg}, while large DL algorithms may require seconds per inference. 
The goal of this study is to enable the use of these algorithms in online and offline data processing tiers, in the context of the LHC experiments' increasing data rates. 
Our approach does this by offloading the computational burden of these algorithms to GPUs while making minimal changes to existing CPU-based workflows.

To achieve this, we move existing work a step further by exploiting the ``as-a-service'' paradigm, in which user access to applications running on remote cloud infrastructure is provided through a thin client interface~\cite{NIST}. 
In this paper, we design a prototypical framework for LHC computing as a service. 
We apply DL algorithms to replace domain-specific algorithms to solve a variety of physics problems through DL inference. 
We then transfer the algorithm to a coprocessor on an independent (local or remote) server and re-configure the existing CPU nodes to communicate with this server through asynchronous and non-blocking inference requests. 
With the inference task offloaded as a request to the server, the CPU is free to perform the rest of the necessary computing within the event.  

Deploying GPUs as a service (GPUaaS) is a natural way to incorporate alternative coprocessors that has several advantages over a direct-connection approach.
In particular, deploying GPUaaS increases hardware cost-effectiveness by reducing the number of GPUs required to achieve the same throughput. 
This is possible because each GPU can service many more CPUs than a direct-connection paradigm would allow.
It is nondisruptive to the existing LHC computing model by offloading the specific algorithms with minimal client-side re-configuration (see Section~\ref{sec:XaaS}).
It facilitates seamless integration and scalability of heterogeneous coprocessors (such as GPUs and FPGAs), as suited for optimal algorithmic performance.
Finally, by exploiting existing open-source, widely-adopted frameworks that have been optimized for fast GPU-based DL, this approach can be adapted quickly to different tasks at the LHC and beyond.

In this paper, we present several examples of integrating GPUaaS into LHC workflows. 
We consider three DL-based algorithms that span a variety of LHC computing applications. 
We integrate these algorithms into both online and offline LHC workflows with GPUaaS and
we benchmark them to evaluate the impact of GPUaaS on the operation of the HLT and the offline computing grid. 
With the goal of optimizing throughput in the high-rate LHC computing environment, we focus on accelerating model inference on coprocessors, as opposed to training. Based on our results, we propose a model for incorporating GPUs and other coprocessors into LHC computing workflows.

The remainder of this paper is organized as follows. 
In Section~\ref{sec:related}, we briefly review related work. 
In Section~\ref{sec:XaaS}, we provide an overview of the current LHC computing model, the as-a-service computing model, and we derive metrics that quantify the cost-effectiveness of coprocessors in LHC workflows. 
Section~\ref{sec:algos} describes the three ML-based algorithms to be deployed.
We describe our configuration of the servers in Google Cloud and LHC data centers and evaluate the limitations of  each site in Section~\ref{sec:perf}. 
We also measure several performance-related quantities relevant for full-scale LHC reconstruction as a service, including hardware throughput, network bottlenecks, and scaling with number of GPUs. 
In Section~\ref{sec:scaling}, we determine the hardware and networking requirements for maximizing the throughput of these algorithms at scale for LHC computing as a service.
Finally, we conclude in Section~\ref{sec:conclusion}.
\section{Related work}
\label{sec:related}
Researchers across HEP have investigated the use of GPUs in event reconstruction in a variety of ways. 
At the LHC, the focus has been on implementations for the HLT where faster computing times lead directly to increased throughput. 
For offline computing at the LHC, GPUs have not been pursued since previous usage of GPUs, through a direct connection to the CPU, would require a larger redesign of the LHC computing grid. This work avoids the need for directly connected GPUs by employing GPUaaS, which provides a method to allow GPUs to be run remotely. 
GPUaaS enables GPU use in both the HLT and offline workflows without a large redesign of the existing LHC computing grid. 
Additionally, we utilize DL algorithms, which allow for the use of existing GPU compiler frameworks to quickly obtain optimized code. 

To perform data analysis on reconstructed objects, DL algorithms are extensively used in HEP.  
In this context, training of DL algorithms is almost exclusively performed on GPUs. 
Frameworks such as GooFit~\cite{schreiner2017goofit}, pyhf~\cite{lukas_heinrich_2020_4110938}, and hepaccelerate~\cite{pata2019processing} exploit GPU acceleration for HEP maximum likelihood fits and data processing applications.

Within the context of online computing, GPUs were first integrated into the 180 compute nodes of the HLT workflow of the ALICE experiment at the LHC to perform charged particle reconstruction.
They were part of the ALICE operations during 2010--2013 and 2015--2018~\cite{ALICE:2019jgt}.
More recently, GPUs are being considered for the HLT of the LHCb experiment~\cite{Aaij:2019zbu,bruch2020realtime}.
Within the CMS experiment, GPUs have been explored for reconstruction of tracks with the pixel detector~\cite{bocci2020heterogeneous} and charged particle reconstruction, through the use of cellular automata~\cite{Funke_2014}, and through the use of an accelerated pattern recognition algorithm~\cite{7581775}. 
Furthermore, algorithms for GPUs and FPGAs have been developed for real time processing of ring imaging Cherenkov detectors for the LHCb HLT~\cite{articlerich,7543150}. 
Beyond the LHC experiments, GPU algorithms have been developed for the trigger readout of the Mu3e experiment~\cite{vomBruch:2017fqw}, and the dark matter experiment NA62~\cite{Ammendola_2018}. 
These algorithms are planned to run in the next round of data taking for each experiment, starting in 2021. 
In all instances, GPUs have been considered in the context of direct connection to CPUs via PCI Express. 

Within the context of offline computing, GPU use in HEP has remained limited.
In neutrino physics, GPUs have been used for simulation of the propagation of Cherenkov light signatures for the IceCube experiment~\cite{DBLP:conf/eScience/ChirkinDKORRSS19}. 
The experiment recently performed a large-scale test of a GPU-only simulation of neutrino signatures, using over 50,000 GPU cores for a period of 20~minutes\cite{sfiligoi2020running,DBLP:conf/eScience/ChirkinDKORRSS19}. 
In this scenario, a large number of cores are used to accelerate a certain component of IceCube simulation.

The offline and online reconstruction software for large LHC experiments consists of several million lines of CPU code. 
Rewriting this code to run on GPUs, for example using CUDA, would be prohibitively costly. In some cases, such as non-parallelizable or transfer-limited operations, it would likely lead to substantially worse timing performance. 
In this paper, we present, for the first time, an alternative model whereby only algorithms with substantial speedups are ported to GPUs, with each GPU serving many CPU nodes. 
We demonstrate that GPUaaS can be integrated within full LHC workflows and can produce significant overall algorithmic speed improvements. 
A similar model for the utilization of CPUs and FPGAs within the LHC workflow was presented in Ref.~\cite{Duarte:2019fta} using the Services for Optimized Network Inference on Coprocessors (SONIC) framework~\cite{SonicSW}.
The study exploited the Microsoft Brainwave service~\cite{configurable-cloud-acceleration} and demonstrated a decrease in deep neural network (DNN) inference time by nearly 3 orders of magnitude when using an FPGA compared to a CPU. 
This paper extends SONIC to support GPUaaS, demonstrating a viable model for fast and nondisruptive integration of GPUs into the LHC workflow.
Outside of the SONIC work described above and recent upgrades in the ALICE software stack~\cite{Buncic:2015ari}, computing as a service has not previously been pursued in HEP.
\section{As-a-service computing for LHC physics}
\label{sec:XaaS}
\begin{figure}[htbp]
\centering
\includegraphics[keepaspectratio, width=0.85\columnwidth]{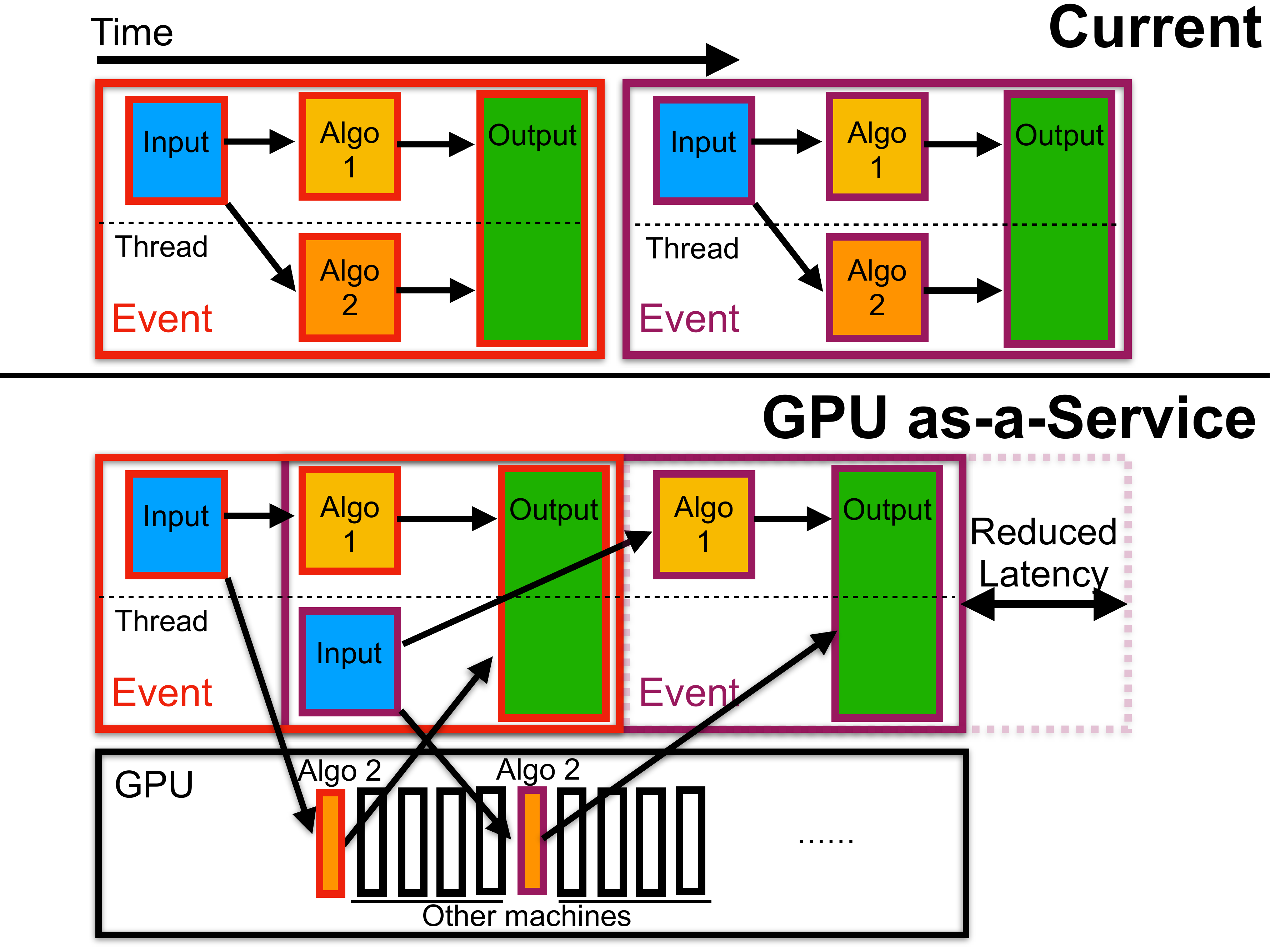}
\caption{Diagram comparing the traditional LHC production model on CPUs (upper) with the GPUaaS approach (lower). 
Each block represents a module within the reconstruction framework. 
For the GPUaaS approach, algorithm 2 is run on the GPU, which allows the processing of the second event (outlined in purple) to run concurrently with the first event (outlined in red). }
\label{fig:gpuaas}
\end{figure}

The current LHC computing model is shown in Figure~\ref{fig:gpuaas}. 
In typical LHC event reconstruction, data is processed sequentially event-by-event, possibly on multiple threads on the CPU.
However, if certain algorithms are significantly accelerated by the use of coprocessors (as shown in Ref.~\cite{Duarte:2019fta}), a modified scenario with coprocessing as a service can be considered.
In this model, a single coprocessor can serve hundreds of CPU processing elements. 
The CPUs are executing numerous different algorithms of the full event reconstruction, whereas the inference server is executing a single algorithm very efficiently. 
To benefit from this type of computing model, there must be a sufficiently large acceleration such that the overhead of offloading this processing onto a separate server does not further increase the reconstruction latency. 
To explain when this is the case, we first review the reconstruction model at the LHC and then discuss how as-a-service computing can be implemented within the LHC reconstruction workflow. 

\subsection{LHC reconstruction}

Detectors at the LHC are general-purpose devices with millions of channels, each of which records information from particles passing through or decaying within it. 
Event reconstruction involves combining these individual signals from different detector channels as optimally as possible to form the set of observable particles, including their energy, momentum, and type, for each event. 
This collection of particles is then used to infer the underlying physics process. 
For example, an event containing a Higgs boson decaying to a bottom quark-antiquark pair can lead to roughly 100 particles and we can use the aggregate properties of these particles to infer the presence of a Higgs boson.
The variety of particles with different signatures in each detector (physics objects) that may be present in any given collision leads to a large number of different reconstruction algorithms that must be run on each event as each physics object typically has its own reconstruction algorithm.
This, in turn, leads to a large codebase that is written entirely for CPUs. 

Parallelization of the reconstruction algorithms that create particles is possible by splitting the reconstruction into separate geometric regions and reconstructing the individual particles within that region. 
Further parallelization is possible through the separate reconstruction within the individual detectors before they are aggregated into particles. 
The current reconstruction aims to exploit possible parallelization by compartmentalizing separate reconstruction algorithms into modules that can be run in parallel. 
No single algorithm dominates the overall computing time, but some fundamental tasks, such as tracking and clustering detector hits to form particles, are the most computationally intensive. 
The potential for parallelization has only partly been realized through standard CPU optimizations, such as auto-vectorization.
In this work, large-scale batching of the reconstruction to allow for algorithm-level parallelism is achieved through the use of DL algorithms on the GPU.



\subsection{As-a-service computing}
To apply DL under the as-a-service paradigm, we choose an algorithm that has a significant speedup when using a GPU. 
We then take this algorithm and set up a GPU inference server using the NVIDIA Triton Inference Server~\cite{Triton}. 
This package uses a custom gRPC-based communication protocol~\cite{gRPC}, and it supports load-balancing between multiple GPUs hosted together in a single server. 
By default, the server queues and processes single-event inference requests with a batch size defined by the client. 
In some cases, the throughput can be increased by aggregating requests from multiple single-event calls, as discussed in Section~\ref{sec:dynbatch}. 
Inference requests can be made for models from various ML frameworks, and multiple models can be loaded on the same server. 
Software frameworks in HEP are typically written in C++. 
The software framework for the CMS experiment, CMSSW~\cite{CMSTDR}, uses task-based multithreading enabled by tbb~\cite{IntelTBB}. 
This facilitates asynchronous, non-blocking calls to external resources using a feature called ExternalWork~\cite{makortelCHEP2019}. 
This is the most efficient way to utilize coprocessors as a service because the CPU running the experiment software can perform other tasks while the service call finishes. 
For this paper, we have taken advantage of these features by extending the CMSSW version of the SONIC software to perform remote gRPC calls to GPUs via the Triton Inference Server. 
In the SONIC approach, only the client code needs to be provided in the software framework. 
This minimizes the maintenance burden, as the client code has just two responsibilities: converting between the experiment and server data formats, and making the call to the server via the chosen communication protocol. 
All the details of the model architecture, any optimizations, and even the choice of coprocessors can be decided on the server without any change in the client. 
This setup enables the modified reconstruction workflow depicted in Figure~\ref{fig:gpuaas}. 

By extending the SONIC framework to handle the gRPC calls utilized by the NVIDIA Triton Inference Server, the new client code uses a standard interface such that the user-developed software to convert between experiment and server data formats remains independent. 
Beyond the specification of the remote server protocol and location within a global configuration file, the user code remains completely intact, and switching between the remote FPGA calls, remote GPU calls, and other local calls are done seamlessly through a configuration file. 

\subsection{Metrics for optimization}




To determine the cost-effectiveness of deploying a given algorithm as a service, we compose a simplified heuristic. 
We assume a computing model similar to those used by LHC experiments, which schedules modules to run during the service request, as in Figure~\ref{fig:gpuaas}. 
We introduce the GPU-to-CPU replacement ratio \Feqgpu to maintain the same throughput:
\begin{equation}
\Feqgpu = \frac{X-S-L}{Y},
\end{equation}
where $X$ is the algorithm processing time on the CPU, $S$ is the overhead time due to input/output packaging in SONIC, and ${L=f(Y+T)}$ is the rescheduling time as a function of the processing time on the GPU $Y$ (which depends on the algorithm, hardware, and batch size), and the packet transfer time $T$. 
For instance, a value of $\Feqgpu=32$ implies that one GPU can replace 32 CPUs at no cost in the overall event throughput. 
While the number of CPUs required to achieve the same throughput decreases due to algorithmic speedups on GPUs, a baseline CPU farm will always be required to perform core event processing and make calls to the GPU. 
A pre-existing CPU farm can serve this baseline requirement. 
The optimal value of $\Feqgpu$ depends on the demands of the system design, as well as the algorithm- and software-dependent values for both $S$ and $L$. 
Time spent in data transfer or queuing on the server plays a small role in total throughput because of the asynchronous, non-blocking call employed in SONIC.

We use \Feqgpu as a guide to contextualize our results for GPU acceleration for each of the different scenarios studied. 
It is derived provided that no substantial bottlenecks are present in the software infrastructure, and further studies will refine this model. 
In the following sections, we explore the GPU speedups utilizing the SONIC framework for algorithms with various \Feqgpu values. 
We discuss the discovered bottlenecks and present a path towards a realistic implementation of GPUaaS at the LHC.  
An interesting potential extension of these studies would be to systematically investigate the relationship between $\Feqgpu$ and throughput and latency.
\section{Algorithms}
\label{sec:algos}
To investigate the scalability of deploying DL as a service for LHC experiments, we study three distinct algorithms.
Together, these algorithms span LHC computing, from low-level tasks of local detector energy reconstruction to high-level tasks of offline object identification.
They also exhibit a range of speedups on coprocessors.
Each algorithm performs as well as a CPU reference algorithm at resolving physical quantities.
We then accelerate these algorithms with GPUaaS in realistic LHC workflows. 
We report the results of these tests in the next sections.

While the emphasis in this paper is on DL algorithms because optimized GPU implementations already exist, many LHC algorithms are currently not ML-based and likely will remain that way in the future.
Nonetheless, many of these tasks have been shown to benefit significantly in computational performance if deployed on coprocessors with custom implementations~\cite{rovere2020clue}. 
The technology we develop for the ML algorithms as a service is flexible and its extension to non-ML algorithms is straightforward.

\subsection{Hadron calorimeter reconstruction}
The simplest algorithm that we study is called Fast Calorimeter Learning (FACILE), a deep neural network consisting of 2,000 parameters. 
This algorithm was trained on a CPU using simulated collisions at the LHC using generator-level information to reconstruct the energy deposited by particles in each cell of the CMS hadron calorimeter (HCAL). 
FACILE uses 15 inputs that contain information about raw charge collected by detector hardware, coordinates of the HCAL channel, and gain of the HCAL channel. 
FACILE consists of batch normalization~\cite{pmlr-v37-ioffe15} and dense layers with rectified linear unit (ReLU) activation functions~\cite{10.5555/3104322.3104425,pmlr-v15-glorot11a}. 
The layers consist of 30, 20, 10, 5, and 3 neurons each, respectively. 
The authors trained the network using a sample size exceeding 1 million HCAL channel events separated into training and testing datasets. 
A mean squared error loss function, batch size of 5,000, and the Adam optimizer~\cite{kingma2017adam} were used in training. 

The HCAL is a core component of LHC experiments and a prototypical subdetector for which to implement ML as-a-service reconstruction for several reasons. 
First, good resolution in the HCAL is important for sensitive measurements in particle physics, such as events with a Higgs boson decaying to bottom quarks. 
We find that local (e.g. HCAL channel energies) and global (e.g. jets and missing transverse momenta) physics objects reconstructed with FACILE have similar resolution to objects reconstructed with the nominal algorithm that does not use ML~\cite{Lawhorn_2019}.
Second, the nominal HCAL reconstruction algorithm in CMS requires $X = 60$\unit{ms} of CPU time, accounting for approximately 15\% of the online computing budget~\cite{ACAT2019}. 
FACILE offers a significant improvement in computing performance when operated as a service by reducing the CPU time (approximating $S+L$) to less than 7 ms, resulting in an estimated $\Feqgpu = 27$, with respect to the nominal algorithm on CPU, which we verify experimentally. 
The remote time (including $Y = 2$\unit{ms} of GPU latency) is reduced by the asynchronous, non-blocking ExternalWork feature employed by SONIC. 
Finally, by exploiting GPU performance for large batch sizes, FACILE offers enhanced physics potential by reconstructing all 16,000 HCAL channels in parallel with little added latency, instead of reconstructing only the highest energy channels.

In terms of physics and computational performance, FACILE is well-suited for both online and offline applications. 
We test it in both settings. 
For instance, we perform a high-bandwidth test designed to emulate, for the first time at scale, a realistic LHC online computing system with coprocessors as a service.
	
\subsection{Electron regression}
DeepCalo is a midsize convolutional neural network (CNN) trained for electron energy regression for the ATLAS detector~\cite{Faye:2019}.
It operates at a higher, more abstract level compared to FACILE since it reconstructs the energy from an entire region of a calorimeter subdetector. 
Compared to nominal techniques~\cite{Aaboud:2018ugz,Aad:2019tso}, DeepCalo improves energy resolution and robustness against pileup, both of which are important for the HL-LHC~\cite{Faye:2019}.
The model is trained on electrons reconstructed from a Monte Carlo simulation of collisions spanning a wide range of energies.
Each collision deposits energy in the electromagnetic calorimeter (ECAL) cells. 
These energy deposits are encoded as a $56\times 11$ pixel image with 4 channels that represents a 2D patch of the detector of width $0.175$ in $\eta$ and $0.270$ in $\phi$.
The 4 channels represent 4 separate layers of the ECAL and each pixel value represents the amount of energy deposited at that location and in that layer in $\eta$ and $\phi$.
Using these images, DeepCalo estimates the energy of the electron.

DeepCalo is composed of 1.8 million parameters. 
The first component of the CNN consists of 5 convolutional blocks. 
The first block performs a $5\times 5$ convolution, followed by batch normalization and a leaky ReLU activation function~\cite{leakyrelu}. 
Each subsequent convolutional block performs a $2\times 2$ maximum pooling, followed by two instances of a sub-block consisting of a $3\times 3$ convolution, batch normalization, and a leaky ReLU activation function. 
The final component of DeepCalo consists of three fully-connected layers, with the last layer producing a prediction for the electron energy. 

In this study, we deploy DeepCalo as a service on GPU coprocessors for offline reconstruction. 
In the offline test, we maximize the event throughput and compare the performance to on-site, CPU-based implementations. 
When deployed as a service, DeepCalo shows significant performance gains by reducing the latency per event from 75 to 1.5\unit{ms} and, when optimized with dynamic batching (as described in Section~\ref{sec:dynbatch}), to 0.1\unit{ms}. 
This yields an estimated $\Feqgpu = 50$ nominally and $750$ after optimization. 


\subsection{Top Quark Tagging}

ResNet-50~\cite{resnet50} is a CNN composed of 23 million parameters, 49 convolutional layers of $7\times 7$, $3\times 3$, and $1\times 1$ convolutions with ``skip connections,'' and 1 fully-connected layer, which predicts 1000 class probabilities for natural images.
In earlier studies~\cite{Duarte:2019fta}, the ResNet-50 CNN architecture was re-purposed to identify events containing top quarks (top quark tagging). 
In addition, CMS has implemented similar CNN-based top quark tagging algorithms for offline reconstructions~\cite{Sirunyan:2020lcu}.
Another study~\cite{Kasieczka_2019} showed that ResNet-50 could be modified to perform top quark tagging with performance rivaling leading ML algorithms. 
Of the three algorithms, ResNet-50 is the most complex and has the longest latency on GPU, and we estimate $\Feqgpu = 150$. 
We choose it to enable benchmarking of a CPU-prohibitive algorithm as a service.  
In particular, ResNet-50 has a CPU latency of the order of seconds, which is prohibitively high for use even in offline reconstruction scenarios. 

In Ref.~\cite{Duarte:2019fta}, we observed a speedup by orders of magnitude by deploying ResNet-50 as a service on FPGA coprocessors. 
In this study, we extend our earlier studies by deploying ResNet-50 as a service on GPU coprocessors in LHC workflows. 
This enables top quark tagging to be performed in offline reconstruction.
ResNet-50 also serves as a prototypical large benchmark algorithm comparable in burden to other major tasks in LHC computing, such as tracking. 
The specifications and GPU utilization of the three algorithms are summarized in Table~\ref{tab:algo-specs}.  

\begin{table*}[htbp]
\centering
\caption{\label{tab:algo-specs}Summary of the specifications of each algorithm, including model parameters, GPU memory usage, and GPU utilization. 
The memory usage and GPU utilization are quoted at the point of maximum GPU throughput. 
For FACILE and DeepCalo, the quantities correspond to an NVIDIA V100 GPU, while for ResNet-50, the quantities are quoted for an NVIDIA T4 GPU. }
\resizebox{\textwidth}{!}{
\begin{tabular}{l|ccccccc}
\hline\noalign{\smallskip}
\multirow{2}{*}{Algorithm} & Batch & 
Architecture &
Trainable & Number & GFLOP & 
GPU memory & 
GPU utilization \\
 & size & type & parameters & of layers & per batch & usage [GB] & [\%]\\
\noalign{\smallskip}\hline\noalign{\smallskip}
FACILE     & 16\,k & Dense & 2\,k & 5 & 0.032 & 1 & 20 \\
DeepCalo   & 5    & Convolutional & 2\,M & 13 & 0.43 & 2.6 & 40 \\
ResNet-50   & 10    & Convolutional & 23\,M & 50 & 39 & 12 & 95 \\
\noalign{\smallskip}\hline
\end{tabular}}
\end{table*}

\section{GPU performance studies}
\label{sec:perf}
For online computing, we integrate FACILE into the HLT, the second tier of CMS data acquisition. 
For offline computing, we consider all three algorithms in stand-alone workflows. 
The client implementation is based on SONIC in a separate CMSSW fork~\cite{SonicSW}. 
We quantify the hardware and networking requirements to run these algorithms as a service in LHC computing. 
To achieve this, we measure the coprocessor throughput (in events processed per second), the number of servers and GPUs required to service a given number of clients (or simultaneous processes), and the network bandwidth limitations (arising from on-premises and external sources) in both LHC computing clusters and on the Google Cloud Platform.

We focus on achieving a hardware-efficient deployment of the algorithms by monitoring server properties and GPU utilization. 
We also measure how throughput scales with the number of GPUs by deploying many GPUs on a single server with a customized Google Kubernetes engine setup (as described in Section~\ref{sec:optimizations}). 
Finally, we investigate various optimizations for the GPUs to further increase the throughput.
Ultimately, in Section~\ref{sec:scaling}, we apply our findings to determine the hardware and networking requirements to perform full-scale LHC computing with coprocessors as a service.

\subsection{Online Computing}
To study the use of coprocessors as a service in online computing, we run the full CMS HLT with local HCAL reconstruction performed by FACILE as a service.
FACILE is particularly well-suited for an online computing application because the algorithm it replaces is responsible for 15\% of the HLT latency per event. 
In this study, the clients are deployed as jobs running single-thread HLT instances on virtual machines in Google Cloud using the HEPCloud framework~\cite{Holzman_2017,altunay2018intelligentlyautomated,Mhashilkar_2019}. 
HEPCloud deploys jobs submitted on batch systems to CPU instances created dynamically at a cloud computing site. 
The jobs are synchronized by adding a waiting period such that each job begins processing information only when all jobs are ready. 
This ensures that all jobs send calls to the GPU server during the same time period, enabling an accurate measurement of GPU and network throughput. 
Since FACILE has a small GPU latency (2 ms) compared to the HLT (500 ms), it proved essential to run on CPUs absent of other jobs for a realistic emulation of the current system of dedicated HLT cores. 
The cloud enabled this by reducing systematic uncertainties arising from shared CPUs on-premises. 
The server was deployed at the same site and consisted of a Google compute instance with either 1 or 4 NVIDIA Tesla V100 GPUs. 
This client-server configuration realistically emulates a fraction of the dedicated HLT CPU farm at CERN with the addition of as-a-service computing. 

\begin{figure}[htbp]
\centering
  \includegraphics[keepaspectratio, width=0.85\columnwidth]{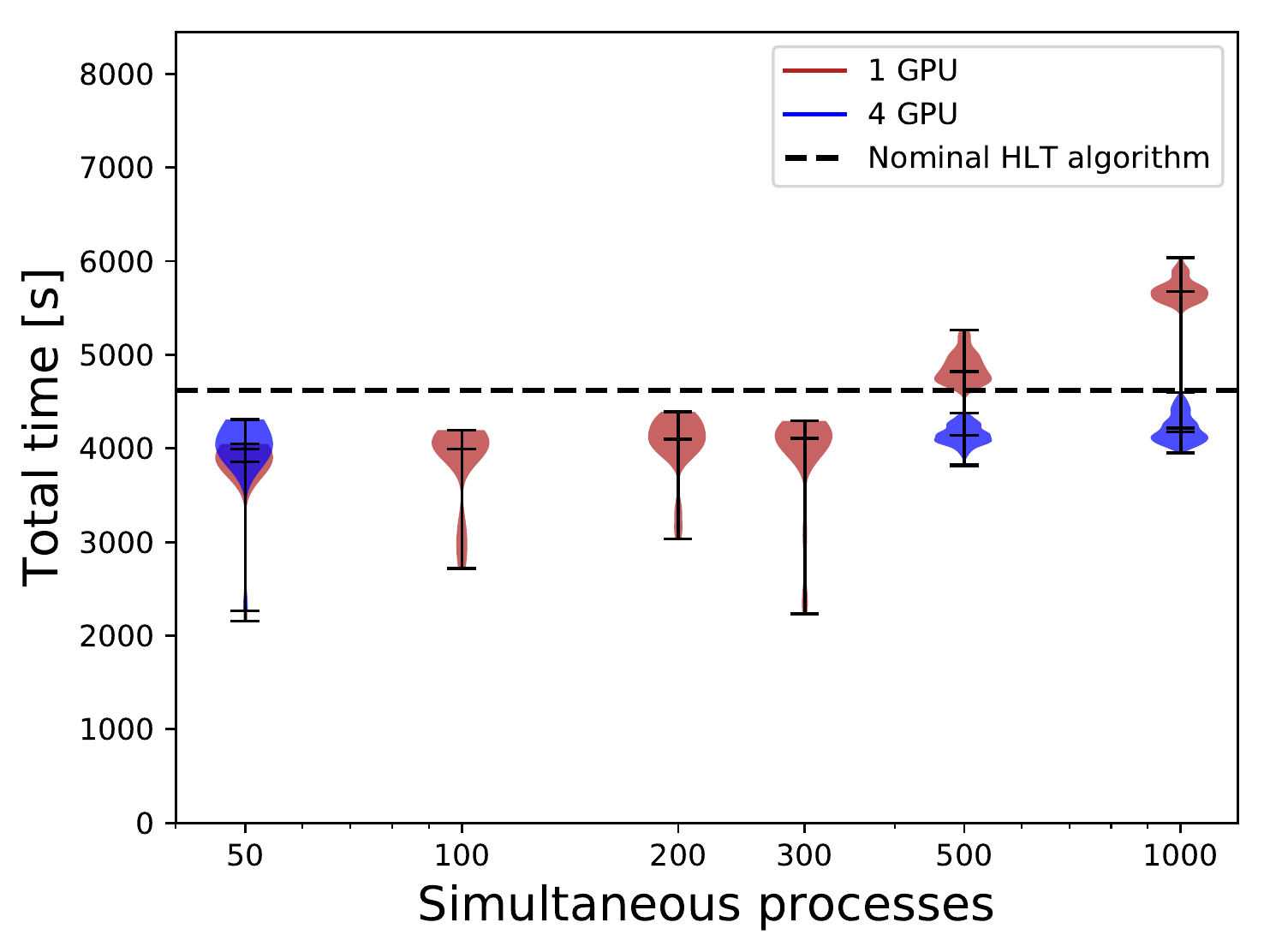}
  \includegraphics[keepaspectratio, width=0.85\columnwidth]{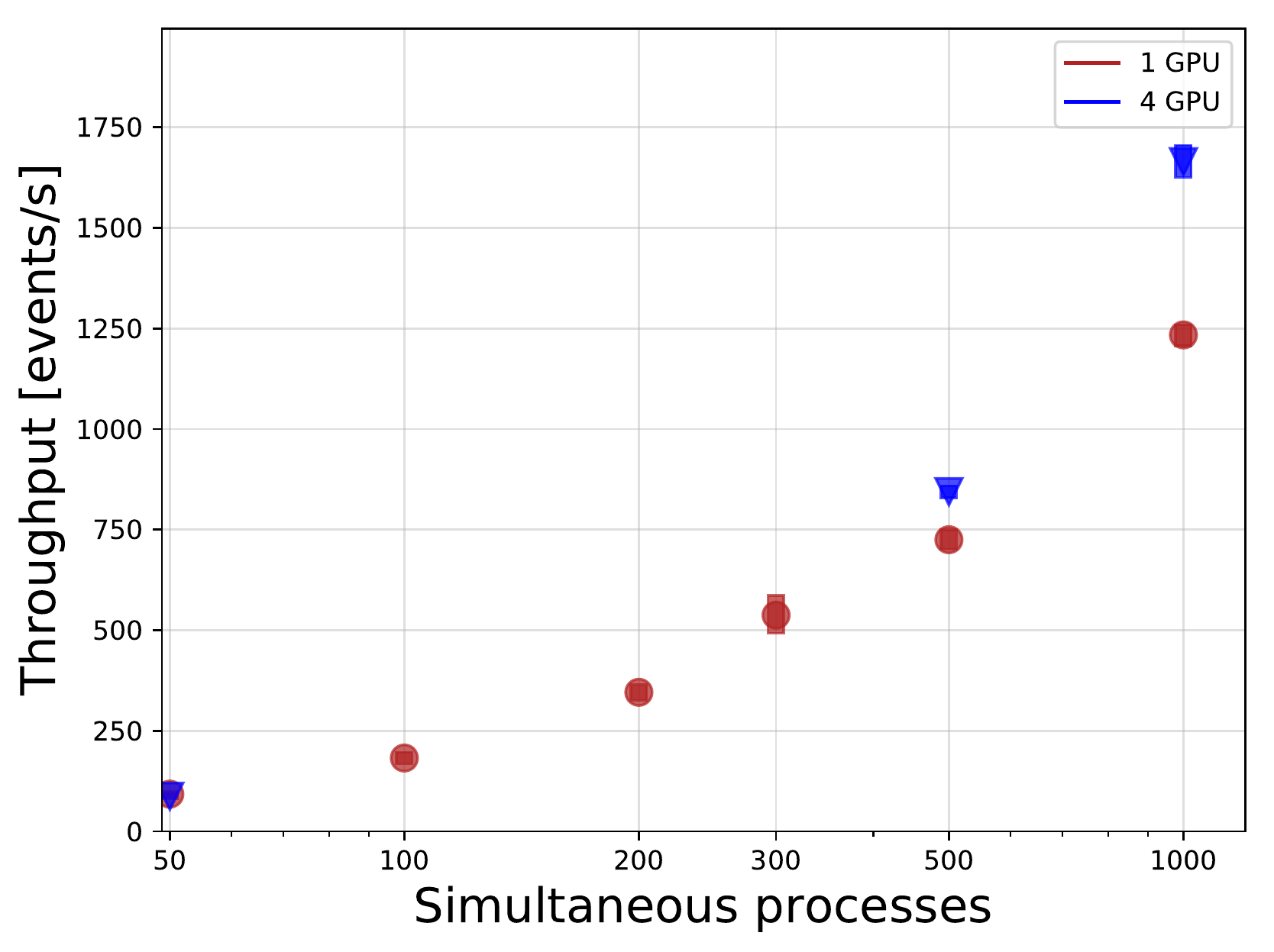}
  \caption{The distribution of the total time (including median and whiskers) to run the high level trigger with the HCAL reconstruction performed with FACILE as a service (upper).
  Servers with 1 and 4 GPUs are shown as red and blue violins. 
  The average time taken to process the same events using the nominal HCAL reconstruction is shown as a dotted black line. 
  High level trigger throughput running FACILE for servers with 1 and 4 GPUs in red and blue markers, respectively (lower). Data between 100-300 simultaneous processes are omitted for the 4 GPU server.}
  \label{fig:hcal_hlt_throughput}
\end{figure}

The results of this test are shown in Figure~\ref{fig:hcal_hlt_throughput}. 
Each client is allotted 7,000 simulated LHC benchmark timing events. 
The timing distribution for the HLT running FACILE as a service is shown in the top panel for servers with 1 or 4 GPUs in red and blue violins. For the HLT tests, FACILE is operated at 4,500 batch size per event; however, the algorithm operates at similar latency with the full HCAL detector (batch size 16,000).  
The average time to run the nominal HLT algorithm locally on the CPU is shown in a dotted black line. 
For fewer than 500 clients, a decrease of approximately 10\% in the total time is observed with FACILE as a service with 1 GPU when compared to the nominal algorithm. 
This largely eliminates the CPU burden of HCAL reconstruction. 
Since the HLT farm at CERN operates under latency restrictions, this demonstrates an opportunity to increase the throughput of the current trigger system by 10\%, or alternatively, partitioning 10\% of the existing machines to be used for other tasks. 
An increase in aggregate HLT latency occurs only above 300 clients for a single GPU, and above 1,000 clients for 4 GPUs. 
This increase represents the point where GPU throughput limitations begin to dominate, indicating that at least 300 HLT instances can be serviced by a single GPU without penalty. 
This slightly exceeds our expectation of 180 HLT instances, based on our computation of $\Feqgpu=27$ divided by the 15\% CPU time fraction, but confirms the overall scaling. We attribute this overperformance to the fact that the actual number of HLT calls for the algorithm is less than the rate at which the algorithm is run.
As a result, we conclude that operating reconstruction as a service is more efficient than having GPUs directly connected to CPUs, since more than 32 cores can be serviced by a single GPU. 
We explore this further by describing a scale design in Section~\ref{sec:scaling}. 
We note that the long tails in the figure are caused by scheduler assignments where fewer jobs are run on certain machines, leading to improved throughput for a small number of jobs, but a negligible effect in overall throughput. 

The HLT throughput with FACILE as a service is shown in the bottom panel of Figure~\ref{fig:hcal_hlt_throughput}. 
For the single GPU server (red circles), the throughput starts demonstrating asymptotic behaviour above 300 simultaneous processes, while for the 4 GPU server (blue triangles), it does not yet asymptote even up to 1,000 simultaneous processes. 

\subsection{Offline Reconstruction}

\begin{figure}[htbp]
\centering

\includegraphics[keepaspectratio, width=0.85\columnwidth]{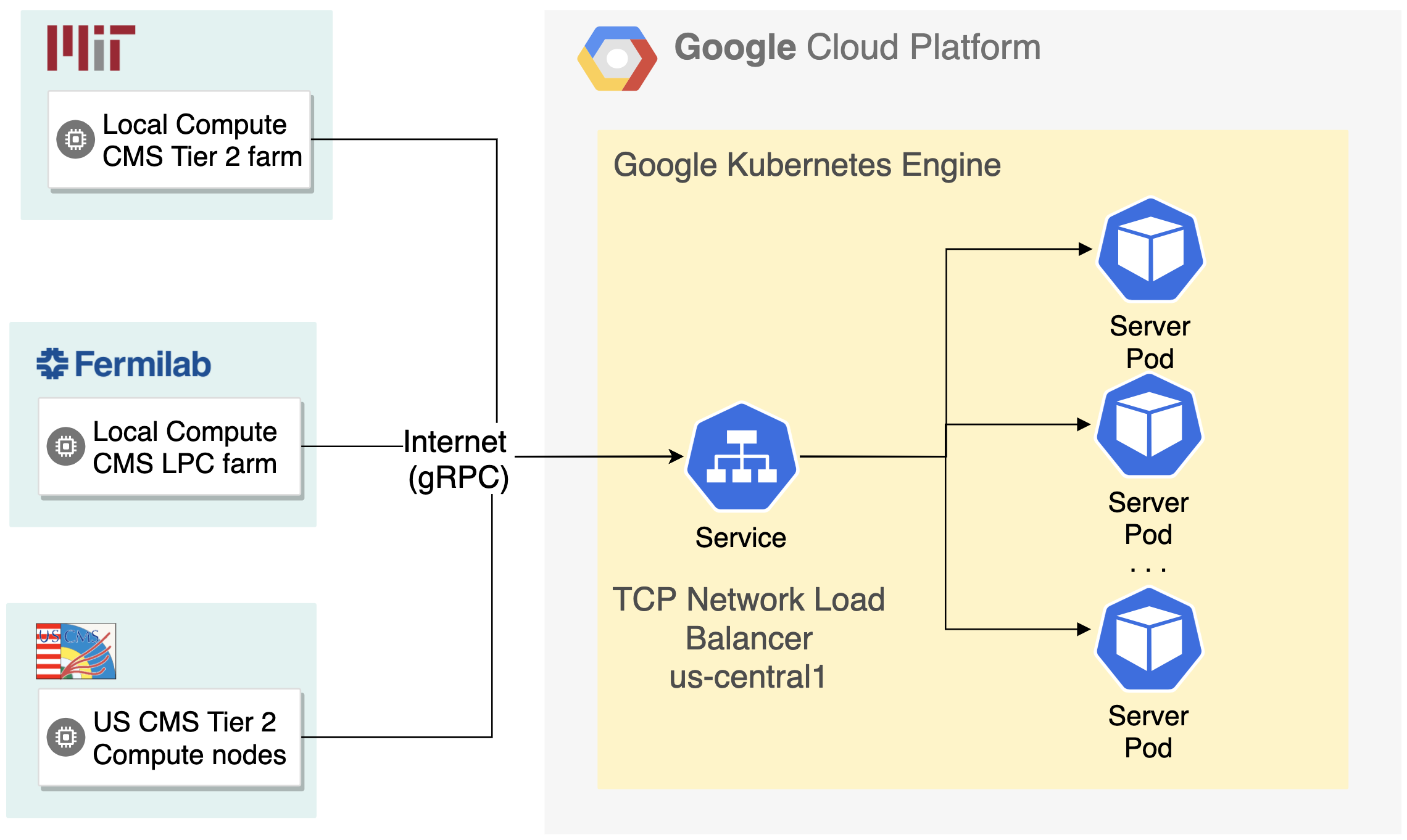}
\caption{Diagram of the architecture for large-scale processing using GPU as a service. 
In this scenario, a GPU server within Google Cloud (right) is used to serve many offline computing centers processing LHC data (left). 
The calls are sent remotely over the internet as gRPC requests.}
\label{fig:gpucloud}
\end{figure}

In the offline computing scenario, a single GPU service can be used by several remote computing clusters at the same time as depicted in Figure~\ref{fig:gpucloud}. 
We investigate the use of FACILE, DeepCalo, and ResNet-50 for LHC offline computing by executing a dedicated workflow process for each model. 
Given the approximately 150,000 CPU cores available for a single LHC experiment and that event reconstruction times are on the order of 30~s per event, our tests assume a benchmark LHC computing throughput of 5,000 events per second, and we estimate the coprocessors necessary to attain this. 
The processing of each model includes realistic input, formatting, and output steps. 
To emulate a realistic global offline computing scenario with CPU workers, as shown in Figure~\ref{fig:gpucloud}, we deploy clients to CPU clusters at MIT and Fermilab. 
These CPUs send gRPC requests over the internet to servers in Google Cloud's \texttt{us-central1-a} zone in Council Bluffs, Iowa. 
While not shown here, we repeated these same tests on-premises going from on-site CPUs to the GPU with Google Cloud and we observed nearly identical throughput saturation and networking effects to the tests observed when going from a remote location to the same GPU within Google Cloud. 
This implies that communication over distance is reliable at the network bandwidths of interest.

\subsubsection{FACILE}

\begin{figure}[htbp]
\centering

  \includegraphics[width=0.85\columnwidth]{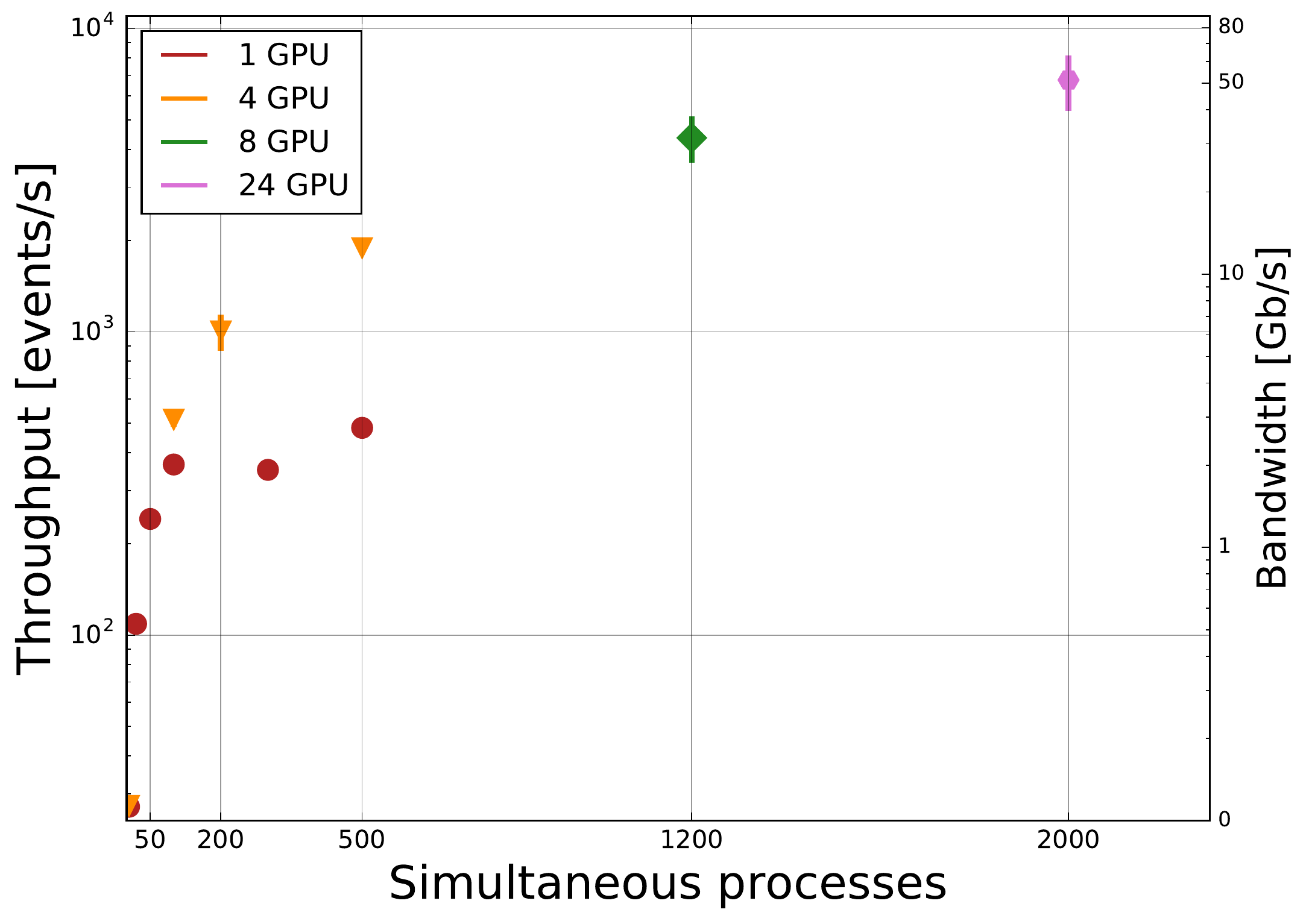}
  \caption{The throughput of FACILE operated as a service in events per second versus the number of simultaneous clients. 
  The marker indicates the mean and the error bars indicate the standard deviation. 
  The throughput tests are repeated with 1, 4, 8, and 24 GPUs connected to the server, represented by red circles, orange triangles, green diamonds, and pink hexagons, respectively. 
  The network bandwidth from the transfer of inputs from the client to the server is shown on the right vertical axis.}
  \label{fig:hcal_offline_throughput}
\end{figure}

The throughput of FACILE as a service is shown for different numbers of clients and GPUs in Figure~\ref{fig:hcal_offline_throughput}. 
A server with a single GPU is found to saturate at a throughput of 500 inferences per second for a V100 GPU. 
This limit is due to the hardware latency and occurs above 50 clients. 
The test is operated at a batch size of 16,000 per event, providing a conservative assumption of HCAL reconstruction requirements.
The V100 GPU is used because it offers a 10--20\% gain over other GPU models. 

As we increase the number of GPUs on the server using a customized Google Kubernetes Engine setup (see \ref{sec:optimization_monitoring}, we find that the throughput scales linearly and with high efficiency. 
Servers with 4 and 8 GPU saturate at approximately 2,000 and 4,000 inferences per second, as shown in Figure~\ref{fig:hcal_offline_throughput}, respectively. 
Therefore, the LHC throughput requirement can be satisfied by a single 10 GPU server. 
This indicates that the Google Kubernetes Engine employed here is an efficient way to increase the throughput. 
The 24 GPU server test with 2,000 clients, in pink, is designed to probe the limit on network bandwidth between the LHC clusters and Google Cloud. 
This test becomes limited by a network bottleneck of unknown origin and we observe a peak bandwidth exceeding 70~Gb/s. 
The number of clients at which saturation occurs also scales with the number of GPUs; for example, the 4 GPU server in orange does not saturate until nearly 500 clients. 
We note that we were not able to plot out the entire throughput distributions due to the expense of each test. 

The throughput is highly sensitive to the server configuration. 
Initially, we deployed a server with a single 4 CPU ingress node handing off the request to nodes with GPUs. 
These tests proved to be limited to a throughput of 1,500 inferences per second (12~Gb/s) regardless of client number, indicating there was a bandwidth limitation at the destination rather than between MIT and Google Cloud. 
As a result, we iteratively reconfigured our server to deploy multiple machines behind a load balancer, as described in Section~\ref{sec:optimization_monitoring}. 

\subsubsection{DeepCalo}

As DeepCalo performs an image classification task, we expect it to be computationally bound rather than bandwidth limited. 
In our studies, we investigate the application of DeepCalo in offline reconstruction, which is throughput limited. 
We evaluate the performance of running DeepCalo as a service by running up to 1,000 clients on-premises on Fermilab's computing cluster and deploying a GPU server in Google cloud. 
We set the batch size to 5 because this is the approximate number of electrons expected per reconstructed collision in a realistic LHC scenario.

\begin{figure}[t]
\centering

  \includegraphics[keepaspectratio, width=0.85\columnwidth]{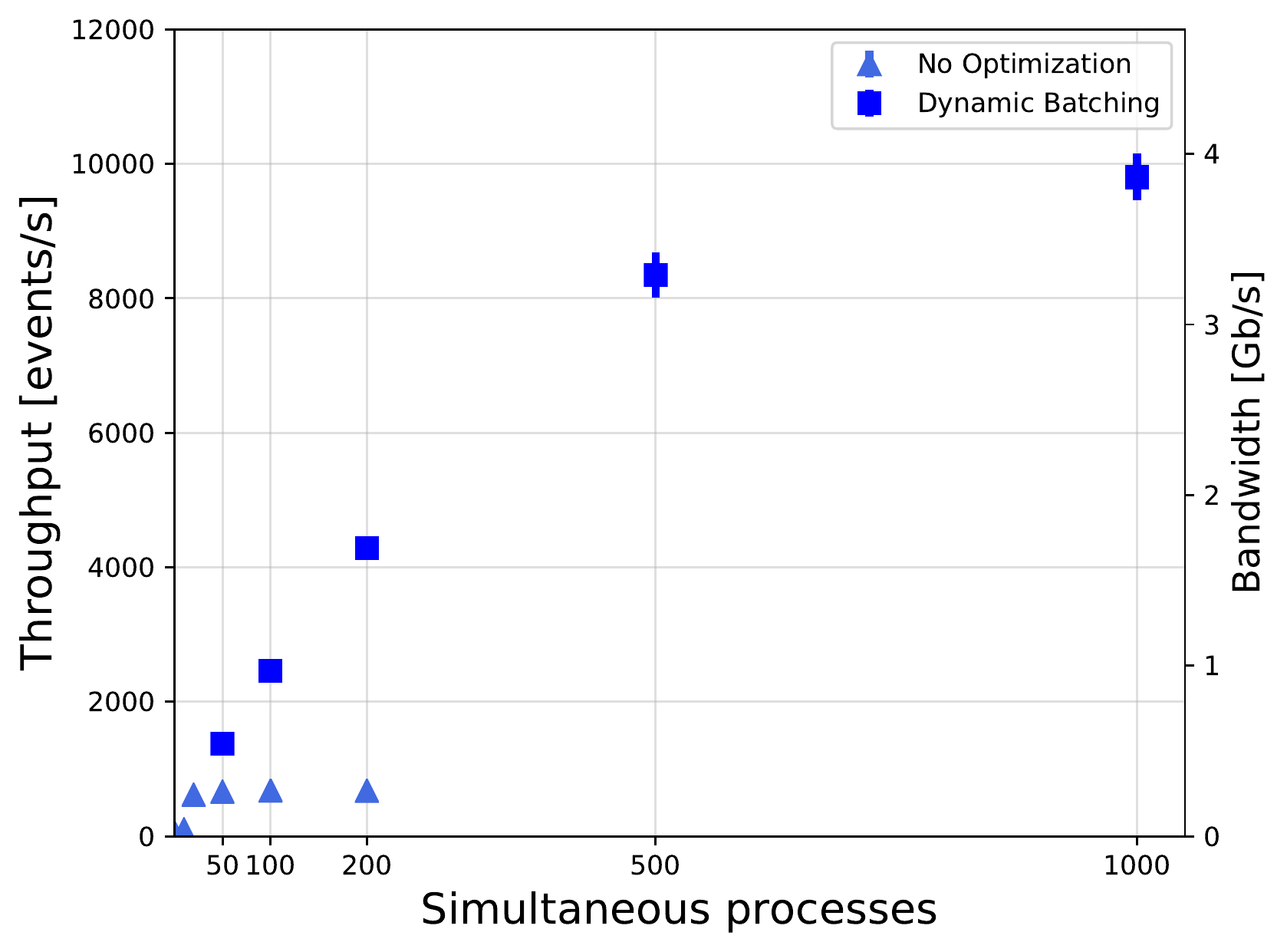}
  \caption{The throughput of DeepCalo as a service on an NVIDIA V100 GPU in events per second versus the number of simultaneous clients. 
  The markers indicate the mean throughput and the error bars indicate the standard deviation. 
  With server side optimizations, the batch size is configured dynamically to prefer a batch of 250 or 500 (as explained in Section~\ref{sec:dynbatch}) and to use five concurrent model executions. 
  The network bandwidth from the transfer of inputs from the client to the server is shown on the right vertical axis.}
  \label{fig:deepcalo_dynamic_throughput}
\end{figure}

We consider the case of a single NVIDIA V100 GPU server deployed on Google Cloud. 
The results are shown in Figure~\ref{fig:deepcalo_dynamic_throughput}. 
For a batch size of 5, the throughput increases rapidly until 20 simultaneous clients, and it saturates at 680 events per second between 20 and 50 simultaneous clients. 
At 200 simultaneous clients, the utilization of the GPU saturates at 45\% and the bandwidth peaks at 270~Mb/s, suggesting that the batch size is limiting GPU utilization. 
Further optimizations to DeepCalo, namely dynamic batching, are discussed in Section~\ref{sec:dynbatch}.

In a previous study, the latency on four Xeon E5-2698 CPUs was found to be 15 ms per electron--or 75 ms for an event of 5 electrons~\cite{Faye:2019}. 
With our baseline GPU performance, we compute 680 events per second or 1.5~ms per event. Combining our GPU result and CPU result, we observe a factor of 50 improvement in the throughput. 

\subsubsection{ResNet-50}
\begin{figure}[htbp]
\centering

  \includegraphics[keepaspectratio, width=0.85\columnwidth]{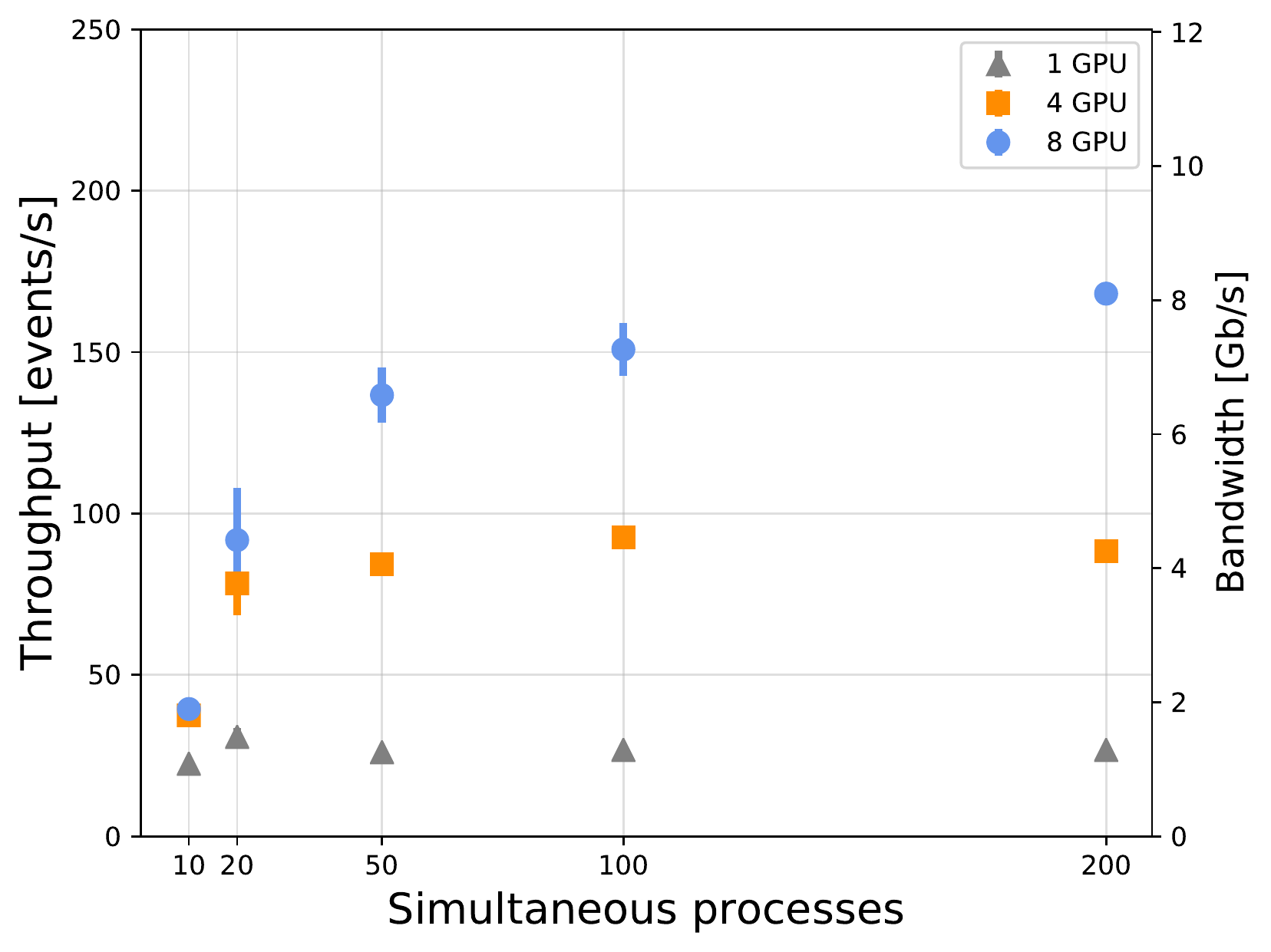}
  \caption{Throughput of ResNet-50 as a service in events per second versus the number of simultaneous clients. 
  The markers indicate the mean throughput and the error bars indicate the standard deviation. 
  The throughput tests are repeated with 1, 4, and 8 GPUs connected to the server, represented by yellow triangles, gold squares, and orange circles, respectively. 
  The network bandwidth from the transfer of inputs from the client to the server is shown on the right vertical axis.}
  \label{fig:resnet50_t4_scaling_gpu_throughput}
\end{figure}

ResNet-50 is deployed as a service with clients on Fermilab's computing cluster and servers with 1, 4, and 8 NVIDIA Tesla T4 GPUs in Google Cloud. 
The throughput obtained using ResNet-50 are shown in Figure~\ref{fig:resnet50_t4_scaling_gpu_throughput}. 
A single GPU server saturates at 25 (batch 10) inferences per second, at about 10 clients. We find that the throughput scales linearly with the number of GPUs, although with approximately 85\% efficiency, slightly lower than for FACILE. 


\subsection{Optimizations}
\label{sec:optimizations}
\subsubsection{Dynamic batching}
\label{sec:dynbatch}
Dynamic batching is a feature of the Triton Inference Server that serves to increase both the throughput and hardware efficiency. 
It performs an added server-side queue of requests from the client until an optimal batch size is reached. 
The performance gains of dynamic batching are also related to the ``instance groups,'' or simultaneous executions of a model. 
This poses an optimization problem between the dynamic batch size and model concurrency. 
The use of dynamic batching is particularly interesting in that it circumvents the LHC computing paradigm of splitting computations on an event-by-event basis. 
Here, multiple events can be processed simultaneously within a single computation, without redesigning the computing model. 
We stress that this type of scheduling is \emph{only} beneficial when GPUs are servicing a large number of parallel processes. 


In the initial DeepCalo measurements, we found that the choice of batch size limited the utilization and throughput of the GPU. 
In our studies, we found that a low number of model execution instances and a high dynamic batch size yielded the best throughput result. Figure~\ref{fig:deepcalo_dynamic_throughput} shows the throughput gains when using dynamic batching. 
At 200 simultaneous clients, the throughput shows no signs of saturation; at this point, the throughput is about 4,200 events per second, representing a factor of 6 improvement. 
When extended to 1,000 simultaneous clients, the throughput reaches 9,800 events per second, representing a factor of 14 gain in throughput, and the GPU utilization increases to 80\%. 
At 1,000 clients, the bandwidth peaks at 3.9~Gb/s, which is roughly the same bandwidth limit observed with FACILE. On the other hand, dynamic batching for FACILE yielded no gain in throughput. 
We conclude that the most significant gains of dynamic batching are found for large models that naturally operate with small batch sizes.

\subsubsection{Server optimization and monitoring}
\label{sec:optimization_monitoring}
We performed tests on many different combinations of computing hardware, which provided us with a deeper understanding of networking limitations within Google Cloud and on-premises data centers. 
Even though the Triton Inference Server does not consume significant CPU power, the number of CPU cores provisioned for the node did have an impact on the maximum ingress bandwidth reached in our early tests.

To scale the GPU throughput flexibly, we deployed a Google Kubernetes Engine cluster for server side workloads. 
The cluster was configured using a Deployment and ReplicaSet~\cite{gcpNetDocs}, which control Pods, groups of one or more containers with shared storage and network and a specification to run the containers~\cite{k8sDocs}, and their resource requests. 
Additionally, a load balancing service was deployed which distributed incoming network traffic among the Pods. 
We implemented Prometheus-based monitoring~\cite{prometheus} of overall system health and inference statistics.
All metrics were visualized through a Grafana~\cite{grafana} instance, also deployed in the same cluster.

We note a Pod’s contents are always co-located and co-scheduled, and run in a shared context within Kubernetes Nodes~\cite{k8sDocs}. 
We kept the Pod to Node ratio at 1:1 throughout the studies, with each Pod running an instance of Triton Inference Server (v20.02-py3) from the NVIDIA Docker repository. 

It can be naively assumed that a small instance \texttt{n1-standard-2} with 2 vCPUs, 7.5\unit{GB} of memory, and different GPU configurations (1, 2, 4, 8) would be able to handle the workload. However, Google Cloud imposes a hard limit on network bandwith per virtual CPU (vCPU). 
After performing several tests, we found that horizontal scaling would allow us to increase our ingress bandwidth since Google Cloud imposes a hard limit on network bandwidth at 2\unit{Gb/s} per vCPU up to a theoretical maximum of 16\unit{Gb/s} for each virtual machine~\cite{gcpNetDocs}.


Given these parameters, the ideal setup for optimizing ingress bandwidth was to provision multiple Pods on 16-vCPU machines with fewer GPUs per Pod. 
For GPU-intensive tests, we took advantage of having a single point of entry through Kubernetes load balancing and provisioning multiple identical Pods, where the sum of the GPUs attached to each Pod is the total GPU requirement.


\subsubsection{Future optimizations}
Throughout these tests, we monitored the GPU utilization. 
ResNet-50 largely saturated GPU utilization, whereas FACILE and DeepCalo used 20\% and 45\% of the GPU, respectively. 
This indicates the throughput is batch limited.
Throughput and GPU utilization can be optimized using dynamic batching for DeepCalo, as described above, but not for FACILE. 
Follow-up studies will investigate optimizations for small models like FACILE.

\section{Scaling} \label{sec:scaling}
We now apply our findings to determine the GPU resources, networking and compute resources required for a modified LHC computing system in which algorithms with large speedups on coprocessors are deployed as a service. 
To assess the amount needed, we compute the required resources for the scaling of the three algorithms in either the HLT or offline computing cases. 
As a benchmark, we use the plateau performance numbers measured per GPU for each of the algorithms. 
We summarize these in Table~\ref{tab:algosummary}. 
As an estimate of the total amount of required resources, we make some assumptions for the typical latency and throughput that we would expect for the online reconstruction and offline reconstruction. 
We emphasize that these numbers are approximate and used here for illustrative purposes.

\begin{table}[htbp]
\caption{Summary of the three algorithms in terms of replacement with respect to a CPU ($\Feqgpu$), inferences per second, individual packet size (Mb), and network bandwidth per GPU (Gb/s). 
These numbers are quoted for the desired batch per event. 
The asterisk ($\ast$) identifies an algorithm using dynamic batching.}
\begin{center}
\begin{tabular}{l|cccc}
\hline\noalign{\smallskip}
\multirow{2}{*}{Algorithm} & \multirow{2}{*}{$\Feqgpu$} & Inf./s & Size/batch & Throughput/GPU  \\
 & & [Hz] & [Mb] & [Gb/s] \\
\noalign{\smallskip}\hline\noalign{\smallskip}
FACILE     & 27  & 500  & 7.7 & 3.9  \\
DeepCalo   & 50  & 680 & 0.4 & 0.3  \\
DeepCalo$^\ast$   & 750  & 9,800 & 0.4 & 3.9  \\
ResNet-50   & 150 & 25   & 48.2 & 1.2  \\
\noalign{\smallskip}
\hline
\end{tabular}
\end{center}
\label{tab:algosummary}
\end{table}

A typical LHC HLT consists of 1,000 servers, each with 32 cores, for a total of 32,000 CPU cores. 
For a 0.5\unit{s} latency, this system would process events  at a rate of 64,000 events per second. 
The goal of the HLT is decide whether the event is sufficiently interesting to preserve for further reconstruction. 
As a consequence, the HLT performs a sequential, tiered reconstruction and filtering of the event and immediately rejects the event if it fails any filter in the sequence in order to prevent further reconstruction~\cite{Perrotta:2015jyu,Donato:2017zlw}. 
With an emulated HLT, we find that a single GPU running FACILE can serve 300--500 different HLT nodes while preserving a 10\% reduction in the per event latency. 
This means that roughly 100 GPUs are needed to serve FACILE for the whole HLT. 
Moreover, if 100 GPUs were added to the system, 10\% or 3,200 CPU cores could be removed from the system. 
We emphasize that FACILE represents only the tip of the iceberg in the use of deep learning in low-level reconstruction at the LHC.
The algorithm uses less than a tenth of the GPU memory, and its GFLOP is less than a tenth of the other algorithms (see Table~\ref{tab:algo-specs}). 
As a result, it can be extended to carry out sophisticated reconstruction tasks (thus merging multiple algorithms) without large increases in latency.

While there is a significant reduction in the number of processing cores, there is an increase in network usage. 
With each GPU, an additional network bandwidth of 3.9\unit{Gb/s} is required. 
A server of 10 GPUs would thus require 40\unit{Gb/s} while simultaneously serving 100 HLT servers (3,200 cores). 
While this bandwidth is large, it is already attainable.
A setup of one 10 GPU server, serving 100 HLT servers, would be a logical design for the system that could be implemented with existing technology.

Lastly, we consider the option of adapting the DeepCalo algorithm or ResNet-50 to run in the HLT with our benchmark batch values. 
We expect that a 6 GPU server with a bandwidth of 24 Gb/s would be sufficient to run DeepCalo for the whole HLT system. 
ResNet-50 with the default batch size of 10 images would require 2,500 GPUs with a total added bandwidth of 2.9\unit{Tb/s}. 
The large number of GPUs and the large bandwidth would require significant and costly modifications to the design of the current HLT, making it impractical. 
However, a ResNet-50 implementation using a batch size of 1, or equivalently an inference rate below the expectation for a batch size of 1 by applying the algorithm only to some events, would result in a comparable number of GPU servers to FACILE. Here, Resnet-50 is only used to identify hadronically decaying top quarks. The rate of objects within an event that have properties that would require  top quark identification is very low. Therefore, batch-1 or lower is more reflective of the true physics use case. 
This conclusion meshes well with the fact that high-energy top quark candidates are relatively rare so it may be reasonable to run a ResNet-50 top quark tagging algorithm once or less per event.

To run FACILE in the existing LHC offline computing system with the approximate throughput of 5,000 reconstructed events per second, a single server of 10 GPUs operating with a bandwidth of 39\unit{Gb/s} would be needed to sustain the full reconstruction load. 
Applying the same scaling for DeepCalo, we find that 1 GPU would be sufficient to run the reconstruction for a whole LHC experiment. 
Lastly, for ResNet-50 at a batch size of 10, a setup of 200 GPUs with 240\unit{Gb/s} bandwidth would be sufficient to support 150,000 CPU cores. 
In this case, with ResNet-50 applied to all reconstructed events, a realistic scenario would consist of 10 separate GPU servers, each running with 24\unit{Gb/s} bandwidth. 
In contrast, for ResNet-50 inference on CPUs with a batch size of 10, the reconstruction time per event would increase by 18\unit{s}. 
This would require a 60\% increase in the computing clusters or an additional 90,000 cores to sustain the same throughput.

In the context of LHC algorithms, DL algorithms are being developed at all levels of the detector reconstruction. 
Algorithms that run on aggregate event properties are comparable in size to ResNet-50, such as jet tagging algorithms~\cite{Qu:2019gqs,Sirunyan:2020lcu,Kasieczka_2019}. 
The full collection of particles in a collision after reconstruction is found to be on the order of 1,000 particles per event~\cite{Sirunyan_2018}. 
Given the number of particles, a computation of the event size ranges from 0.5--2.5\unit{Mb}, which is less than the size of a single event request performed in tests with FACILE. 
Consequently, we observe that data rates and throughput for future LHC algorithms are comparable to that of the results presented with FACILE. 
Therefore, at no added complexity in networking or design,  the framework presented in this paper can be extended for algorithms designed for the future particle reconstruction at the LHC. 

\section{Conclusion}
\label{sec:conclusion}
We have demonstrated a core framework that enables the use of deep learning (DL) as a service with direct applications to the processing of LHC data.
Our framework, Services for Optimized Network Inference on Coprocessors (SONIC), utilizes gRPC to perform asynchronous, non-blocking calls to a GPU server. 
Our server infrastructure can address both small and large scale use cases. 
With our infrastructure, we have tested three algorithms that span a large range of DL model complexities and batch sizes. 
Together, these algorithms serve as benchmarks for a wide array of LHC reconstruction tasks. 
In each case, we have measured the algorithm throughput and demonstrated comparable throughput for as-a-service computing both remotely and on-site. 
We fully integrated a DL algorithm called Fast Calorimeter Learning (FACILE) for LHC reconstruction in the high-level trigger (HLT), the second tier of the LHC data processing and filter, and we found that this algorithm can lead to a 10\% overall reduction in the computing demands. 
This latency reduction is almost equivalent to removing the hadron calorimeter reconstruction latency from the HLT entirely, and it matches the expected optimal performance when performing standalone algorithmic tests. 
Finally, we demonstrated the use of FACILE, DeepCalo, and ResNet-50 for offline reconstruction.
A server implementation in the cloud was found to operate at data rates and inference times comparable to the demands set by LHC offline reconstruction. 
This is a concrete validation of the SONIC framework, demonstrating the viability of coprocessors as a service on representative scales for LHC computing. 

While our focus was on accelerating DL algorithms with GPUs, this work can be applied to any algorithm that can be implemented on a GPU and appropriately integrated into a GPU server.
This work is largely agnostic to the hardware and software implementations of the algorithm and can be adapted for other types of coprocessors and other scientific experiments. 
As DL accelerator tools are constantly evolving and improving, we expect the speedups observed in this paper to become even larger.


From our studies, we delineated certain considerations for designing an optimal system with GPUs as a service (GPUaaS) for the LHC.
In particular, an optimized scheduling framework is needed to ensure that remote operations of algorithms incur minimal losses in performance. 
Additionally, sufficient bandwidth is needed to ensure that the full performance of the accelerator servers can be achieved for both remote and local as-a-service operation. 
Finally, both a load balancer and an optimized and flexible server infrastructure are needed to ensure robust operation. In this paper, we have demonstrated that all of these demands can be met with existing resources. 

In the context of physics performance, our results lead to direct performance improvements that can be implemented immediately in the LHC computing model. 
In particular, we found that: (1) DL inference as a service can enable a significant increase in event throughput, (2) algorithms with complexities not previously attainable can be operated in the LHC reconstruction workflow, (3) optimized versions of algorithms can be implemented without disrupting the existing computing model, and (4) simultaneous multi-event processing is achievable in the reconstruction workflow. 
Concurrently with these studies, an extensive suite of new DL techniques for LHC reconstruction have been developed~\cite{collaboration2019deep,Metodiev:2017vrx,Nachman:2020lpy,Andreassen:2020nkr,Collins:2018epr,Collins:2019jip,Farina:2018fyg,Heimel:2018mkt,Qasim:2019otl,Farrell:2018cjr,ExaTrkX,Qu:2019gqs,Moreno:2019bmu,Moreno:2019neq,Choma:2020cry,Bogatskiy:2020tje,Kieseler:2020wcq,Mikuni:2020wpr,Pata:2021oez}, which explore new architectures such as equivariant neural networks, graph neural networks, and attention-based transformers. 
The current work will enable the integration of DL algorithms, including these, into the LHC computing model in a seamless and computationally efficient way. 
Our work can also be extended by adapting to the unique computing challenges that emerge from the next generation of DL algorithms. 
In the future, we plan to investigate the requirements and constraints at high performance computing (HPC) centers, in order to leverage their resources for reliable local operations of a large number of GPUs. 
This work can also help drive preparation towards a computing model for future high-luminosity LHC running, where larger rates and numbers of channels will further stretch computing capacity.

We would like to stress that this work represents the beginning of developments in coprocessor computing both at the LHC and other large scale experiments. 
This work and related studies are encouraging for physicists in other fields, such as neutrino physics~\cite{Wang:2020fjr}, gravitational wave detection, and astrophysics, to pursue a similar computing model. 
As a consequence, we believe that this work may lead to a paradigm shift in the scientific computing model, enabling us to meet the enormous scientific computing demands in the next decade. 

\section*{Acknowledgments}
P.~H., and D.~R. are supported by NSF grants \#1934700,  \#1931469, and the IRIS-HEP grant \#1836650. 
J.~K. is supported by NSF grant \#190444 and NSERC PGS D. 
Cloud credits for this study were provided by Internet2 managed Exploring Cloud to accelerate Science (NSF grant \#190444). 
Additional support on networking was provided by Doug Burton, Dan Speck, and Aaron Strong of the Burwood group. 
Also, we thank Emma Levett from MIT for additional networking support. 
P.~H. would like to thank Matthew Harris for discussion on GCP networking capabilities.  
M.~A.~F, B.~H., T.~K., M.~L., K.~P., N.~T. are supported by Fermi Research Alliance, LLC under Contract No. DE-AC02-07CH11359 with the U.S. Department of Energy (DOE), Office of Science, Office of High Energy Physics.
K.~P. is partially supported by the High Velocity Artificial Intelligence grant as part of the DOE High Energy Physics Computational HEP sessions program.
J.~D. is supported by the DOE, Office of Science, Office of High Energy Physics Early Career Research program under Award No. DE-SC0021187.

\section*{References}
\bibliographystyle{lucas_unsrt} 
\bibliography{references2}

\providecommand{\href}[2]{#2}\begingroup\raggedright\begin{thebibliography}{10}%
\makeatletter
\providecommand{\hrefCMSnoop }[0]{\@secondoftwo}%
\makeatother
\providecommand{\doi}{\texttt{doi:}\begingroup \urlstyle{tt}\Url}

\bibitem{Evans:2008zzb}
\hrefCMSnoop {}{L.~Evans and P.~Bryant, ``{LHC Machine}'',} \textit{ JINST}
  \textbf{ 3} (2008) S08001,
  \href{http://dx.doi.org/10.1088/1748-0221/3/08/S08001}{\doi{10.1088/1748-0221/3/08/S08001}}.

\bibitem{Boyd:2020qox}
\hrefCMSnoop {}{J.~Boyd, ``{LHC Run-2 and Future Prospects}'',} in \textit{
  {2019 European School of High-Energy Physics}}.
\newblock 2020.
\newblock
  \href{http://www.arXiv.org/abs/2001.04370}{\texttt{arXiv:2001.04370}}.

\bibitem{Gerber:2019jlx}
\hrefCMSnoop {}{C.~E. Gerber, ``{LHC Highlights and Prospects}'',} in \textit{
  {10th CERN\textendash{}Latin-American School of High-Energy Physics}}.
\newblock 2019.
\newblock
  \href{http://www.arXiv.org/abs/1909.10919}{\texttt{arXiv:1909.10919}}.

\bibitem{Dainese:2019rgk}
\hrefCMSnoop {}{{ATLAS and CMS Collaborations}, ``{Report on the Physics at the
  HL-LHC, and Perspectives for the HE-LHC}'',} \textit{ CERN Yellow Reports:
  Monographs} \textbf{ 7/2019} (2019)
  \href{http://dx.doi.org/10.23731/CYRM-2019-007}{\doi{10.23731/CYRM-2019-007}},
  \href{http://www.arXiv.org/abs/1902.10229}{\texttt{arXiv:1902.10229}}.

\bibitem{Alimena:2019zri}
\hrefCMSnoop {}{J.~Alimena {et~al.}, ``{Searching for long-lived particles
  beyond the Standard Model at the Large Hadron Collider}'',} \textit{ J. Phys.
  G} \textbf{ 47} (2020) 090501,
  \href{http://dx.doi.org/10.1088/1361-6471/ab4574}{\doi{10.1088/1361-6471/ab4574}},
  \href{http://www.arXiv.org/abs/1903.04497}{\texttt{arXiv:1903.04497}}.

\bibitem{Abercrombie:2015wmb}
\hrefCMSnoop {}{D.~Abercrombie {et~al.}, ``{Dark Matter Benchmark Models for
  Early LHC Run-2 Searches: Report of the ATLAS/CMS Dark Matter Forum}'',}
  \textit{ Phys. Dark Univ.} \textbf{ 27} (2020) 100371,
  \href{http://dx.doi.org/10.1016/j.dark.2019.100371}{\doi{10.1016/j.dark.2019.100371}},
  \href{http://www.arXiv.org/abs/1507.00966}{\texttt{arXiv:1507.00966}}.

\bibitem{:2012gk}
\hrefCMSnoop {}{{ATLAS} Collaboration, ``Observation of a new particle in the
  search for the standard model {Higgs} boson with the {ATLAS} detector at the
  {LHC}'',} \textit{ Phys. Lett. B} \textbf{ 716} (2012) 1,
  \href{http://dx.doi.org/10.1016/j.physletb.2012.08.020}{\doi{10.1016/j.physletb.2012.08.020}},
  \href{http://www.arXiv.org/abs/1207.7214}{\texttt{arXiv:1207.7214}}.

\bibitem{:2012gu}
\hrefCMSnoop {}{{CMS} Collaboration, ``Observation of a new boson at a mass of
  125 {GeV} with the {CMS} experiment at the {LHC}'',} \textit{ Phys. Lett. B}
  \textbf{ 716} (2012) 30,
  \href{http://dx.doi.org/10.1016/j.physletb.2012.08.021}{\doi{10.1016/j.physletb.2012.08.021}},
  \href{http://www.arXiv.org/abs/1207.7235}{\texttt{arXiv:1207.7235}}.

\bibitem{Chatrchyan:2013lba}
\hrefCMSnoop {}{{CMS} Collaboration, ``{Observation of a new boson with mass
  near 125 GeV in pp collisions at $\sqrt{s}$ = 7 and 8 TeV}'',} \textit{ JHEP}
  \textbf{ 06} (2013) 081,
  \href{http://dx.doi.org/10.1007/JHEP06(2013)081}{\doi{10.1007/JHEP06(2013)081}},
\href{http://www.arXiv.org/abs/1303.4571}{\texttt{arXiv:1303.4571}}.

\bibitem{Sirunyan:2017jix}
\hrefCMSnoop {}{{CMS} Collaboration, ``{Search for new physics in final states
  with an energetic jet or a hadronically decaying $W$ or $Z$ boson and
  transverse momentum imbalance at $\sqrt{s}=13\text{ }\text{
  }\mathrm{TeV}$}'',} \textit{ Phys. Rev. D} \textbf{ 97} (2018) 092005,
  \href{http://dx.doi.org/10.1103/PhysRevD.97.092005}{\doi{10.1103/PhysRevD.97.092005}},
  \href{http://www.arXiv.org/abs/1712.02345}{\texttt{arXiv:1712.02345}}.

\bibitem{Sirunyan:2019vgj}
\hrefCMSnoop {}{{CMS} Collaboration, ``{Search for high mass dijet resonances
  with a new background prediction method in proton-proton collisions at
  $\sqrt{s} =$ 13 TeV}'',} \textit{ JHEP} \textbf{ 05} (2020) 033,
  \href{http://dx.doi.org/10.1007/JHEP05(2020)033}{\doi{10.1007/JHEP05(2020)033}},
  \href{http://www.arXiv.org/abs/1911.03947}{\texttt{arXiv:1911.03947}}.

\bibitem{Sirunyan:2019vxa}
\hrefCMSnoop {}{{CMS} Collaboration, ``{Search for low mass vector resonances
  decaying into quark-antiquark pairs in proton-proton collisions at
  $\sqrt{s}=$ 13 TeV}'',} \textit{ Phys. Rev. D} \textbf{ 100} (2019) 112007,
  \href{http://dx.doi.org/10.1103/PhysRevD.100.112007}{\doi{10.1103/PhysRevD.100.112007}},
  \href{http://www.arXiv.org/abs/1909.04114}{\texttt{arXiv:1909.04114}}.

\bibitem{alex2016dark}
\hrefCMSnoop {}{J.~Alexander {et~al.}, ``Dark sectors 2016 workshop:
  {Community} report'',} in \textit{ Dark Sectors 2016 Workshop}.
\newblock 2016.
\newblock
  \href{http://www.arXiv.org/abs/1608.08632}{\texttt{arXiv:1608.08632}}.

\bibitem{Albrecht_2019}
\hrefCMSnoop {}{{HEP Software Foundation} Collaboration, ``A roadmap for {HEP}
  software and computing {R\&D} for the 2020s'',} \textit{ Comput. Softw. Big
  Sci.} (2019) 7,
  \href{http://dx.doi.org/10.1007/s41781-018-0018-8}{\doi{10.1007/s41781-018-0018-8}},
  \href{http://www.arXiv.org/abs/1712.06982}{\texttt{arXiv:1712.06982}}.

\bibitem{octwiki}
\href
  {https://twiki.cern.ch/twiki/bin/view/CMSPublic/CMSOfflineComputingResults}{{CMS}
  Collaboration, ``{CMS} offline and computing public results'',} 2020.
\newblock \url
  {https://twiki.cern.ch/twiki/bin/view/CMSPublic/CMSOfflineComputingResults}.

\bibitem{ATLASCompute}
\href
  {https://twiki.cern.ch/twiki/bin/view/AtlasPublic/ComputingandSoftwarePublicResults}{{ATLAS
  Collaboration}, ``{ATLAS} computing and software public results'',} 2020.
\newblock \url
  {https://twiki.cern.ch/twiki/bin/view/AtlasPublic/ComputingandSoftwarePublicResults}.

\bibitem{HS06}
\href {https://w3.hepix.org/benchmarking.html}{{HEPiX Benchmarking Working
  Group}, ``{HEP-SPEC06}'',} 2017.
\newblock \url {https://w3.hepix.org/benchmarking.html}.

\bibitem{dennard}
R.~H. {Dennard}\hrefCMSnoop {}{ {et~al.}, ``Design of ion-implanted {MOSFET}'s
  with very small physical dimensions'',} \textit{ IEEE J. Solid-State
  Circuits} \textbf{ 9} (1974) 256,
  \href{http://dx.doi.org/10.1109/JSSC.1974.1050511}{\doi{10.1109/JSSC.1974.1050511}}.

\bibitem{breakdown}
H.~Esmaeilzadeh\hrefCMSnoop {}{ {et~al.}, ``Dark silicon and the end of
  multicore scaling'',} in \textit{ Proceedings of the 38th Annual
  International Symposium on Computer Architecture}, ISCA ’11, p.~365.
\newblock ACM, New York, NY, USA, 2011.
\newblock
  \href{http://dx.doi.org/10.1145/2000064.2000108}{\doi{10.1145/2000064.2000108}}.

\bibitem{Duarte:2019fta}
\hrefCMSnoop {}{J.~Duarte {et~al.}, ``{FPGA-accelerated machine learning
  inference as a service for particle physics computing}'',} \textit{ Comput.
  Softw. Big Sci.} \textbf{ 3} (2019) 13,
  \href{http://dx.doi.org/10.1007/s41781-019-0027-2}{\doi{10.1007/s41781-019-0027-2}},
\href{http://www.arXiv.org/abs/1904.08986}{\texttt{arXiv:1904.08986}}.

\bibitem{Guest_2018}
\hrefCMSnoop {}{D.~Guest, K.~Cranmer, and D.~Whiteson, ``Deep learning and its
  application to {LHC} physics'',} \textit{ Annu. Rev. Nucl. Part. Sci.}
  \textbf{ 68} (2018) 161,
  \href{http://dx.doi.org/10.1146/annurev-nucl-101917-021019}{\doi{10.1146/annurev-nucl-101917-021019}},
  \href{http://www.arXiv.org/abs/1806.11484}{\texttt{arXiv:1806.11484}}.

\bibitem{Albertsson:2018maf}
\hrefCMSnoop {}{K.~Albertsson {et~al.}, ``{Machine Learning in High Energy
  Physics Community White Paper}'',} \textit{ J. Phys. Conf. Ser.} \textbf{
  1085} (2018) 022008,
  \href{http://dx.doi.org/10.1088/1742-6596/1085/2/022008}{\doi{10.1088/1742-6596/1085/2/022008}},
  \href{http://www.arXiv.org/abs/1807.02876}{\texttt{arXiv:1807.02876}}.

\bibitem{Bourilkov:2019yoi}
\hrefCMSnoop {}{D.~Bourilkov, ``{Machine and Deep Learning Applications in
  Particle Physics}'',} \textit{ Int. J. Mod. Phys. A} \textbf{ 34} (2020)
  1930019,
  \href{http://dx.doi.org/10.1142/S0217751X19300199}{\doi{10.1142/S0217751X19300199}},
  \href{http://www.arXiv.org/abs/1912.08245}{\texttt{arXiv:1912.08245}}.

\bibitem{Larkoski:2017jix}
\hrefCMSnoop {}{A.~J. Larkoski, I.~Moult, and B.~Nachman, ``{Jet Substructure
  at the Large Hadron Collider: A Review of Recent Advances in Theory and
  Machine Learning}'',} \textit{ Phys. Rept.} \textbf{ 841} (2020) 1,
  \href{http://dx.doi.org/10.1016/j.physrep.2019.11.001}{\doi{10.1016/j.physrep.2019.11.001}},
\href{http://www.arXiv.org/abs/1709.04464}{\texttt{arXiv:1709.04464}}.

\bibitem{Qasim:2019otl}
\hrefCMSnoop {}{S.~R. Qasim, J.~Kieseler, Y.~Iiyama, and M.~Pierini,
  ``{Learning representations of irregular particle-detector geometry with
  distance-weighted graph networks}'',} \textit{ Eur. Phys. J. C} \textbf{ 79}
  (2019) 608,
  \href{http://dx.doi.org/10.1140/epjc/s10052-019-7113-9}{\doi{10.1140/epjc/s10052-019-7113-9}},
\href{http://www.arXiv.org/abs/1902.07987}{\texttt{arXiv:1902.07987}}.

\bibitem{Belayneh:2019vyx}
\hrefCMSnoop {}{D.~Belayneh {et~al.}, ``{Calorimetry with deep learning:
  particle simulation and reconstruction for collider physics}'',} \textit{
  Eur. Phys. J. C} \textbf{ 80} (2020) 688,
  \href{http://dx.doi.org/10.1140/epjc/s10052-020-8251-9}{\doi{10.1140/epjc/s10052-020-8251-9}},
  \href{http://www.arXiv.org/abs/1912.06794}{\texttt{arXiv:1912.06794}}.

\bibitem{Komiske:2016rsd}
\hrefCMSnoop {}{P.~T. Komiske, E.~M. Metodiev, and M.~D. Schwartz, ``{Deep
  learning in color: towards automated quark/gluon jet discrimination}'',}
  \textit{ JHEP} \textbf{ 01} (2017) 110,
  \href{http://dx.doi.org/10.1007/JHEP01(2017)110}{\doi{10.1007/JHEP01(2017)110}},
  \href{http://www.arXiv.org/abs/1612.01551}{\texttt{arXiv:1612.01551}}.

\bibitem{ATL-PHYS-PUB-2017-017}
\href {http://cds.cern.ch/record/2275641}{{ATLAS} Collaboration, ``{Quark
  versus Gluon Jet Tagging Using Jet Images with the ATLAS Detector}'',} ATLAS
  Public Note ATL-PHYS-PUB-2017-017, 2017.

\bibitem{Andrews:2019faz}
M.~Andrews\hrefCMSnoop {}{ {et~al.}, ``{End-to-end jet classification of quarks
  and gluons with the CMS Open Data}'',} \textit{ Nucl. Instrum. Methods Phys.
  Res. A} \textbf{ 977} (2020) 164304,
  \href{http://dx.doi.org/10.1016/j.nima.2020.164304}{\doi{10.1016/j.nima.2020.164304}},
  \href{http://www.arXiv.org/abs/1902.08276}{\texttt{arXiv:1902.08276}}.

\bibitem{WLCGmtg}
\href {https://indico.cern.ch/event/739897/}{{WLCG Grid Deployment Board},
  ``Benchmarking'',} 2019.
\newblock \url {https://indico.cern.ch/event/739897/}.

\bibitem{NIST}
\hrefCMSnoop {}{P.~Mell and T.~Grance, ``{The NIST Definition of Cloud
  Computing}'',} NIST Special Publication SP 800-145, 2011.
\newblock
  \href{http://dx.doi.org/10.6028/NIST.SP.800-145}{\doi{10.6028/NIST.SP.800-145}}.

\bibitem{schreiner2017goofit}
H.~Schreiner\hrefCMSnoop {}{ {et~al.}, ``{GooFit 2.0}'',} \textit{ J. Phys.
  Conf. Ser.} \textbf{ 1085} (2018) 042014,
  \href{http://dx.doi.org/10.1088/1742-6596/1085/4/042014}{\doi{10.1088/1742-6596/1085/4/042014}},
  \href{http://www.arXiv.org/abs/1710.08826}{\texttt{arXiv:1710.08826}}.

\bibitem{lukas_heinrich_2020_4110938}
\hrefCMSnoop {}{L.~Heinrich, M.~Feickert, and G.~Stark, ``scikit-hep/pyhf:
  v0.5.3'',} 10, 2020.
\newblock
  \href{http://dx.doi.org/10.5281/zenodo.4110938}{\doi{10.5281/zenodo.4110938}}.

\bibitem{pata2019processing}
\hrefCMSnoop {}{J.~Pata and M.~Spiropulu, ``Processing columnar collider data
  with {GPU}-accelerated kernels'',}
  \href{http://www.arXiv.org/abs/1906.06242}{\texttt{arXiv:1906.06242}}.

\bibitem{ALICE:2019jgt}
\hrefCMSnoop {}{{ALICE} Collaboration, ``Real-time data processing in the
  {ALICE} high level trigger at the {LHC}'',} \textit{ Comput. Phys. Commun.}
  \textbf{ 242} (2019) 25,
  \href{http://dx.doi.org/10.1016/j.cpc.2019.04.011}{\doi{10.1016/j.cpc.2019.04.011}},
  \href{http://www.arXiv.org/abs/1812.08036}{\texttt{arXiv:1812.08036}}.

\bibitem{Aaij:2019zbu}
\hrefCMSnoop {}{R.~Aaij {et~al.}, ``{Allen}: {A} high level trigger on {GPUs}
  for {LHCb}'',} \textit{ Comput. Softw. Big Sci.} \textbf{ 4} (2020) 7,
  \href{http://dx.doi.org/10.1007/s41781-020-00039-7}{\doi{10.1007/s41781-020-00039-7}},
  \href{http://www.arXiv.org/abs/1912.09161}{\texttt{arXiv:1912.09161}}.

\bibitem{bruch2020realtime}
\hrefCMSnoop {}{D.~Vom~Bruch, ``Real-time data processing with {GPUs} in high
  energy physics'',} \textit{ JINST} \textbf{ 15} (2020) C06010,
  \href{http://dx.doi.org/10.1088/1748-0221/15/06/C06010}{\doi{10.1088/1748-0221/15/06/C06010}},
  \href{http://www.arXiv.org/abs/2003.11491}{\texttt{arXiv:2003.11491}}.

\bibitem{bocci2020heterogeneous}
A.~Bocci\hrefCMSnoop {}{ {et~al.}, ``{Heterogeneous reconstruction of tracks
  and primary vertices with the CMS pixel tracker}'',} \textit{ Front. Big
  Data} \textbf{ 3} (2020) 49,
  \href{http://dx.doi.org/10.3389/fdata.2020.601728}{\doi{10.3389/fdata.2020.601728}},
  \href{http://www.arXiv.org/abs/2008.13461}{\texttt{arXiv:2008.13461}}.

\bibitem{Funke_2014}
D.~Funke\hrefCMSnoop {}{ {et~al.}, ``Parallel track reconstruction in {CMS}
  using the cellular automaton approach'',} \textit{ J. Phys. Conf. Ser.}
  \textbf{ 513} (2014) 052010,
  \href{http://dx.doi.org/10.1088/1742-6596/513/5/052010}{\doi{10.1088/1742-6596/513/5/052010}}.

\bibitem{7581775}
F.~{Pantaleo}\hrefCMSnoop {}{ {et~al.}, ``Development of a {phase-II} track
  trigger based on {GPUs} for the {CMS} experiment'',} in \textit{ 2015 IEEE
  Nuclear Science Symposium and Medical Imaging Conference (NSS/MIC)},
  p.~7581775.
\newblock 2016.
\newblock
  \href{http://dx.doi.org/10.1109/NSSMIC.2015.7581775}{\doi{10.1109/NSSMIC.2015.7581775}}.

\bibitem{articlerich}
\hrefCMSnoop {}{G.~Lamanna, ``Almagest, a new trackless ring finding
  algorithm'',} \textit{ Nucl. Instr. Methods Phys. Res. A} \textbf{ 766}
  (2014) 241,
  \href{http://dx.doi.org/10.1016/j.nima.2014.05.073}{\doi{10.1016/j.nima.2014.05.073}}.

\bibitem{7543150}
\hrefCMSnoop {}{C.~F{\"{a}}erber, R.~Schwemmer, J.~Machen, and N.~Neufeld,
  ``Particle identification on an {FPGA} accelerated compute platform for the
  {LHCb} upgrade'',} \textit{ IEEE Trans. Nucl. Sci.} \textbf{ 64} (2017) 1994,
  \href{http://dx.doi.org/10.1109/TNS.2017.2715900}{\doi{10.1109/TNS.2017.2715900}}.

\bibitem{vomBruch:2017fqw}
\hrefCMSnoop {}{{Mu3e} Collaboration, ``Online data reduction using track and
  vertex reconstruction on {GPUs} for the {Mu3e} experiment'',} \textit{ Eur.
  Phys. J. Web Conf.} \textbf{ 150} (2017) 00013,
  \href{http://dx.doi.org/10.1051/epjconf/201715000013}{\doi{10.1051/epjconf/201715000013}}.

\bibitem{Ammendola_2018}
R.~Ammendola\hrefCMSnoop {}{ {et~al.}, ``Real-time heterogeneous stream
  processing with {NaNet} in the {NA}62 experiment'',} \textit{ J. Phys. Conf.
  Ser.} \textbf{ 1085} (2018) 032022,
  \href{http://dx.doi.org/10.1088/1742-6596/1085/3/032022}{\doi{10.1088/1742-6596/1085/3/032022}}.

\bibitem{DBLP:conf/eScience/ChirkinDKORRSS19}
D.~Chirkin\hrefCMSnoop {}{ {et~al.}, ``Photon propagation using {GPUs} by the
  {IceCube} neutrino observatory'',} in \textit{ 15th International Conference
  on eScience, eScience 2019, San Diego, CA, USA, September 24-27, 2019},
  p.~388.
\newblock {IEEE}, 2019.
\newblock
  \href{http://dx.doi.org/10.1109/eScience.2019.00050}{\doi{10.1109/eScience.2019.00050}}.

\bibitem{sfiligoi2020running}
\hrefCMSnoop {}{I.~Sfiligoi, F.~Wuerthwein, B.~Riedel, and D.~Schultz,
  ``Running a pre-exascale, geographically distributed, multi-cloud scientific
  simulation'',} in \textit{ High Performance Computing}, p.~23.
\newblock 2020.
\newblock
  \href{http://www.arXiv.org/abs/2002.06667}{\texttt{arXiv:2002.06667}}.
\newblock
  \href{http://dx.doi.org/10.1007/978-3-030-50743-5_2}{\doi{10.1007/978-3-030-50743-5_2}}.

\bibitem{SonicSW}
\hrefCMSnoop {}{K.~Pedro, ``{SonicCMS}''.} [software] version v5.1.0 (accessed
  2020-04-22) \url{https://github.com/fastmachinelearning/SonicCMS}, 2020.

\bibitem{configurable-cloud-acceleration}
A.~Caulfield\hrefCMSnoop {}{ {et~al.}, ``A cloud-scale acceleration
  architecture'',} in \textit{ 2016 49th Annual IEEE/ACM International
  Symposium on Microarchitecture (MICRO)}, p.~1.
\newblock IEEE, 2016.
\newblock
  \href{http://dx.doi.org/10.1109/MICRO.2016.7783710}{\doi{10.1109/MICRO.2016.7783710}}.

\bibitem{Buncic:2015ari}
\href {https://cds.cern.ch/record/2011297}{{{ALICE}} Collaboration, ``Technical
  design report for the upgrade of the online-offline computing system'',}
  {ALICE Technical Design Report}, 2015.

\bibitem{Triton}
\hrefCMSnoop {}{{NVIDIA}, ``{Triton Inference Server}''.} [software] version
  v1.8.0 (accessed 2020-02-17)
  \url{https://docs.nvidia.com/deeplearning/sdk/triton-inference-server-guide/docs/index.html},
  2019.

\bibitem{gRPC}
\hrefCMSnoop {}{{Google}, ``{gRPC}''.} [software] version v1.19.0 (accessed
  2020-02-17) \url{https://grpc.io/}, 2018.

\bibitem{CMSTDR}
\href {http://cds.cern.ch/record/922757}{{CMS Collaboration}, ``{{CMS} physics:
  technical design report {Volume} 1: {Detector} performance and software}'',}
  {CMS Technical Design Report} CERN-LHCC-2006-001, 2006.

\bibitem{IntelTBB}
\href {https://www.threadingbuildingblocks.org}{{Intel}, ``{Thread Building
  Blocks}''.} [software] version 2018\_U1 (accessed 2019-02-11)
  \url{https://www.threadingbuildingblocks.org}, 2018.
\newblock 2018\_U1 (accessed 2019-02-11). \url
  {https://www.threadingbuildingblocks.org}.

\bibitem{makortelCHEP2019}
A.~Bocci\hrefCMSnoop {}{ {et~al.}, ``{Bringing heterogeneity to the CMS
  software framework}'',} \textit{ EPJ Web Conf.} \textbf{ 245} (2020) 05009,
  \href{http://dx.doi.org/10.1051/epjconf/202024505009}{\doi{10.1051/epjconf/202024505009}},
  \href{http://www.arXiv.org/abs/2004.04334}{\texttt{arXiv:2004.04334}}.

\bibitem{rovere2020clue}
M.~Rovere\hrefCMSnoop {}{ {et~al.}, ``{CLUE}: {A} fast parallel clustering
  algorithm for high granularity calorimeters in high energy physics'',}
  \textit{ Front. Big Data} \textbf{ 3} (2020) 41,
  \href{http://dx.doi.org/10.3389/fdata.2020.591315}{\doi{10.3389/fdata.2020.591315}},
  \href{http://www.arXiv.org/abs/2001.09761}{\texttt{arXiv:2001.09761}}.

\bibitem{pmlr-v37-ioffe15}
\href {http://proceedings.mlr.press/v37/ioffe15.html}{S.~Ioffe and C.~Szegedy,
  ``Batch normalization: Accelerating deep network training by reducing
  internal covariate shift'',} in \textit{ Proceedings of the 32nd
  International Conference on Machine Learning}, F.~Bach and D.~Blei, eds.,
  volume~37 of \textit{ Proceedings of Machine Learning Research}, p.~448.
\newblock PMLR, Lille, France, 2015.
\newblock
  \href{http://www.arXiv.org/abs/1502.03167}{\texttt{arXiv:1502.03167}}.

\bibitem{10.5555/3104322.3104425}
\hrefCMSnoop {}{V.~Nair and G.~E. Hinton, ``Rectified linear units improve
  restricted {Boltzmann} machines'',} in \textit{ Proceedings of the 27th
  International Conference on International Conference on Machine Learning},
  ICML'10, p.~807.
\newblock Omnipress, Madison, WI, USA, 2010.

\bibitem{pmlr-v15-glorot11a}
\href {http://proceedings.mlr.press/v15/glorot11a.html}{X.~Glorot, A.~Bordes,
  and Y.~Bengio, ``Deep sparse rectifier neural networks'',} in \textit{
  Proceedings of the 14th International Conference on Artificial Intelligence
  and Statistics (AISTATS)}, G.~Gordon, D.~Dunson, and M.~Dud\'{i}k, eds.,
  volume~15, p.~315.
\newblock JMLR Workshop and Conference Proceedings, Fort Lauderdale, FL, USA,
  2011.

\bibitem{kingma2017adam}
\hrefCMSnoop {}{D.~P. Kingma and J.~Ba, ``Adam: A method for stochastic
  optimization'',} in \textit{ 3rd International Conference on Learning
  Representations, {ICLR} 2015, Conference Track Proceedings}, Y.~Bengio and
  Y.~LeCun, eds.
\newblock 2015.
\newblock \href{http://www.arXiv.org/abs/1412.6980}{\texttt{arXiv:1412.6980}}.

\bibitem{Lawhorn_2019}
\hrefCMSnoop {}{J.~Lawhorn, ``New method of out-of-time energy subtraction for
  the {CMS} hadronic calorimeter'',} \textit{ J. Phys. Conf. Ser.} \textbf{
  1162} (2019) 012036,
  \href{http://dx.doi.org/10.1088/1742-6596/1162/1/012036}{\doi{10.1088/1742-6596/1162/1/012036}}.

\bibitem{ACAT2019}
\hrefCMSnoop {}{A.~Massironi, V.~Khristenko, and M.~DAlfonso, ``{Heterogeneous
  computing for the local reconstruction algorithms of the CMS
  calorimeters}'',} \textit{ J. Phys. Conf. Ser.} \textbf{ 1525} (2020) 012040,
  \href{http://dx.doi.org/10.1088/1742-6596/1525/1/012040}{\doi{10.1088/1742-6596/1525/1/012040}}.

\bibitem{Faye:2019}
\hrefCMSnoop {}{F.~Faye, ``Energy reconstruction of electrons and photons using
  convolutional neural networks'',} Master's thesis, University of Copenhagen,
  2019.

\bibitem{Aaboud:2018ugz}
\hrefCMSnoop {}{{ATLAS} Collaboration, ``{Electron and photon energy
  calibration with the ATLAS detector using 2015\textendash{}2016 LHC
  proton-proton collision data}'',} \textit{ JINST} \textbf{ 14} (2019) P03017,
  \href{http://dx.doi.org/10.1088/1748-0221/14/03/P03017}{\doi{10.1088/1748-0221/14/03/P03017}},
  \href{http://www.arXiv.org/abs/1812.03848}{\texttt{arXiv:1812.03848}}.

\bibitem{Aad:2019tso}
\hrefCMSnoop {}{{ATLAS} Collaboration, ``{Electron and photon performance
  measurements with the ATLAS detector using the 2015\textendash{}2017 LHC
  proton-proton collision data}'',} \textit{ JINST} \textbf{ 14} (2019) P12006,
  \href{http://dx.doi.org/10.1088/1748-0221/14/12/P12006}{\doi{10.1088/1748-0221/14/12/P12006}},
  \href{http://www.arXiv.org/abs/1908.00005}{\texttt{arXiv:1908.00005}}.

\bibitem{leakyrelu}
\href
  {https://sites.google.com/site/deeplearningicml2013/relu_hybrid_icml2013_final.pdf}{A.~Maas,
  A.~Jannun, and A.~Ng, ``Rectifier nonlinearities improve neural network
  acoustic models'',} in \textit{ 30th International Conference on Machine
  Learning (ICML), Workshop on Deep Learning for Audio, Speech, and Language
  Processing (WDLASL)}.
\newblock 2013.

\bibitem{resnet50}
\hrefCMSnoop {}{K.~He, X.~Zhang, S.~Ren, and J.~Sun, ``Deep residual learning
  for image recognition'',} in \textit{ 2016 IEEE Conference on Computer Vision
  and Pattern Recognition (CVPR)}, p.~770.
\newblock IEEE, 2016.
\newblock
  \href{http://www.arXiv.org/abs/1512.03385}{\texttt{arXiv:1512.03385}}.
\newblock
  \href{http://dx.doi.org/10.1109/CVPR.2016.90}{\doi{10.1109/CVPR.2016.90}}.

\bibitem{Sirunyan:2020lcu}
\hrefCMSnoop {}{{CMS} Collaboration, ``{Identification of heavy, energetic,
  hadronically decaying particles using machine-learning techniques}'',}
  \textit{ JINST} \textbf{ 15} (2020) P06005,
  \href{http://dx.doi.org/10.1088/1748-0221/15/06/P06005}{\doi{10.1088/1748-0221/15/06/P06005}},
  \href{http://www.arXiv.org/abs/2004.08262}{\texttt{arXiv:2004.08262}}.

\bibitem{Kasieczka_2019}
\hrefCMSnoop {}{A.~Butter {et~al.}, ``The machine learning landscape of top
  taggers'',} \textit{ SciPost Phys.} \textbf{ 7} (2019) 014,
  \href{http://dx.doi.org/10.21468/SciPostPhys.7.1.014}{\doi{10.21468/SciPostPhys.7.1.014}},
  \href{http://www.arXiv.org/abs/1902.09914}{\texttt{arXiv:1902.09914}}.

\bibitem{Holzman_2017}
B.~Holzman\hrefCMSnoop {}{ {et~al.}, ``{HEPCloud}, a new paradigm for {HEP}
  facilities: {CMS} {Amazon Web Services} investigation'',} \textit{ Comput.
  Softw. Big Sci.} \textbf{ 1} (2017) 1,
  \href{http://dx.doi.org/10.1007/s41781-017-0001-9}{\doi{10.1007/s41781-017-0001-9}},
  \href{http://www.arXiv.org/abs/1710.00100}{\texttt{arXiv:1710.00100}}.

\bibitem{altunay2018intelligentlyautomated}
M.~Altunay\hrefCMSnoop {}{ {et~al.}, ``Intelligently-automated facilities
  expansion with the {HEPCloud} decision engine'',} in \textit{ 2018 18th
  IEEE/ACM International Symposium on Cluster, Cloud and Grid Computing
  (CCGRID)}, p.~352.
\newblock 2018.
\newblock
  \href{http://www.arXiv.org/abs/1806.03224}{\texttt{arXiv:1806.03224}}.
\newblock
  \href{http://dx.doi.org/10.1109/CCGRID.2018.00053}{\doi{10.1109/CCGRID.2018.00053}}.

\bibitem{Mhashilkar_2019}
P.~Mhashilkar\hrefCMSnoop {}{ {et~al.}, ``{HEPCloud}, an elastic hybrid {HEP}
  facility using an intelligent decision support system'',} \textit{ Eur. Phys.
  J. Web Conf.} \textbf{ 214} (2019) 03060,
  \href{http://dx.doi.org/10.1051/epjconf/201921403060}{\doi{10.1051/epjconf/201921403060}}.

\bibitem{gcpNetDocs}
\href {https://cloud.google.com/compute/docs/machine-types}{{Google LLC},
  ``{Compute Engine Documentation - Concepts - Virtual machine instances -
  Machine types}'',} 2020.
\newblock \url {https://cloud.google.com/compute/docs/machine-types}.

\bibitem{k8sDocs}
\href {https://kubernetes.io/docs/concepts/workloads/pods/pod/}{{Kubernetes
  Authors}, ``{Documentation - Concepts - Workloads - Pods - Pods}'',} 2020.
\newblock \url {https://kubernetes.io/docs/concepts/workloads/pods/pod/}.

\bibitem{prometheus}
\href {https://prometheus.io/}{{Prometheus Authors}, ``Prometheus'',} 2020.
\newblock \url {https://prometheus.io/}.

\bibitem{grafana}
\href {https://grafana.com/}{{Grafana Labs}, ``Grafana'',} 2020.
\newblock \url {https://grafana.com/}.

\bibitem{Perrotta:2015jyu}
\hrefCMSnoop {}{A.~Perrotta, ``{Performance of the CMS High Level Trigger}'',}
  \textit{ J. Phys. Conf. Ser.} \textbf{ 664} (2015) 082044,
  \href{http://dx.doi.org/10.1088/1742-6596/664/8/082044}{\doi{10.1088/1742-6596/664/8/082044}}.

\bibitem{Donato:2017zlw}
\hrefCMSnoop {}{{CMS} Collaboration, ``{CMS Trigger Performance}'',} \textit{
  Eur. Phys. J. Web Conf.} \textbf{ 182} (2018) 02037,
  \href{http://dx.doi.org/10.1051/epjconf/201818202037}{\doi{10.1051/epjconf/201818202037}}.

\bibitem{Qu:2019gqs}
\hrefCMSnoop {}{H.~Qu and L.~Gouskos, ``{ParticleNet}: Jet tagging via particle
  clouds'',} \textit{ Phys. Rev. D} \textbf{ 101} (2020) 056019,
  \href{http://dx.doi.org/10.1103/PhysRevD.101.056019}{\doi{10.1103/PhysRevD.101.056019}},
  \href{http://www.arXiv.org/abs/1902.08570}{\texttt{arXiv:1902.08570}}.

\bibitem{Sirunyan_2018}
\hrefCMSnoop {}{{CMS} Collaboration, ``Measurement of charged particle spectra
  in minimum-bias events from proton–proton collisions at $\sqrt{s}=13\,\text
  {TeV} $'',} \textit{ Eur. Phys. J. C} \textbf{ 78} (2018)
  \href{http://dx.doi.org/10.1140/epjc/s10052-018-6144-y}{\doi{10.1140/epjc/s10052-018-6144-y}},
  \href{http://www.arXiv.org/abs/1806.11245}{\texttt{arXiv:1806.11245}}.

\bibitem{collaboration2019deep}
\hrefCMSnoop {}{{CMS} Collaboration, ``{A deep neural network to search for new
  long-lived particles decaying to jets}'',} \textit{ Mach. Learn.: Sci.
  Technol.} \textbf{ 1} (2020) 035012,
  \href{http://dx.doi.org/10.1088/2632-2153/ab9023}{\doi{10.1088/2632-2153/ab9023}},
  \href{http://www.arXiv.org/abs/1912.12238}{\texttt{arXiv:1912.12238}}.

\bibitem{Metodiev:2017vrx}
\hrefCMSnoop {}{E.~M. Metodiev, B.~Nachman, and J.~Thaler, ``{Classification
  without labels: Learning from mixed samples in high energy physics}'',}
  \textit{ JHEP} \textbf{ 10} (2017) 174,
  \href{http://dx.doi.org/10.1007/JHEP10(2017)174}{\doi{10.1007/JHEP10(2017)174}},
\href{http://www.arXiv.org/abs/1708.02949}{\texttt{arXiv:1708.02949}}.

\bibitem{Nachman:2020lpy}
\hrefCMSnoop {}{B.~Nachman and D.~Shih, ``Anomaly detection with density
  estimation'',} \textit{ Phys. Rev. D} \textbf{ 101} (2020) 075042,
  \href{http://dx.doi.org/10.1103/PhysRevD.101.075042}{\doi{10.1103/PhysRevD.101.075042}},
  \href{http://www.arXiv.org/abs/2001.04990}{\texttt{arXiv:2001.04990}}.

\bibitem{Andreassen:2020nkr}
\hrefCMSnoop {}{A.~Andreassen, B.~Nachman, and D.~Shih, ``Simulation assisted
  likelihood-free anomaly detection'',} \textit{ Phys. Rev. D} \textbf{ 101}
  (2020) 095004,
  \href{http://dx.doi.org/10.1103/PhysRevD.101.095004}{\doi{10.1103/PhysRevD.101.095004}},
  \href{http://www.arXiv.org/abs/2001.05001}{\texttt{arXiv:2001.05001}}.

\bibitem{Collins:2018epr}
\hrefCMSnoop {}{J.~H. Collins, K.~Howe, and B.~Nachman, ``Anomaly detection for
  resonant new physics with machine learning'',} \textit{ Phys. Rev. Lett.}
  \textbf{ 121} (2018) 241803,
  \href{http://dx.doi.org/10.1103/PhysRevLett.121.241803}{\doi{10.1103/PhysRevLett.121.241803}},
\href{http://www.arXiv.org/abs/1805.02664}{\texttt{arXiv:1805.02664}}.

\bibitem{Collins:2019jip}
\hrefCMSnoop {}{J.~H. Collins, K.~Howe, and B.~Nachman, ``Extending the search
  for new resonances with machine learning'',} \textit{ Phys. Rev. D} \textbf{
  99} (2019) 014038,
  \href{http://dx.doi.org/10.1103/PhysRevD.99.014038}{\doi{10.1103/PhysRevD.99.014038}},
\href{http://www.arXiv.org/abs/1902.02634}{\texttt{arXiv:1902.02634}}.

\bibitem{Farina:2018fyg}
\hrefCMSnoop {}{M.~Farina, Y.~Nakai, and D.~Shih, ``Searching for new physics
  with deep autoencoders'',} \textit{ Phys. Rev. D} \textbf{ 101} (2020)
  075021,
  \href{http://dx.doi.org/10.1103/PhysRevD.101.075021}{\doi{10.1103/PhysRevD.101.075021}},
  \href{http://www.arXiv.org/abs/1808.08992}{\texttt{arXiv:1808.08992}}.

\bibitem{Heimel:2018mkt}
\hrefCMSnoop {}{T.~Heimel, G.~Kasieczka, T.~Plehn, and J.~M. Thompson, ``{QCD
  or What?}'',} \textit{ SciPost Phys.} \textbf{ 6} (2019) 030,
  \href{http://dx.doi.org/10.21468/SciPostPhys.6.3.030}{\doi{10.21468/SciPostPhys.6.3.030}},
\href{http://www.arXiv.org/abs/1808.08979}{\texttt{arXiv:1808.08979}}.

\bibitem{Farrell:2018cjr}
\hrefCMSnoop {}{S.~Farrell {et~al.}, ``{Novel deep learning methods for track
  reconstruction}'',} in \textit{ {4th International Workshop Connecting The
  Dots 2018 (CTD2018)}}.
\newblock 2018.
\newblock
\href{http://www.arXiv.org/abs/1810.06111}{\texttt{arXiv:1810.06111}}.
\newblock

\bibitem{ExaTrkX}
\href
  {https://ml4physicalsciences.github.io/files/NeurIPS_ML4PS_2019_83.pdf}{X.~Ju
  {et~al.}, ``{Graph Neural Networks for Particle Reconstruction in High Energy
  Physics Detectors}'',} in \textit{ {Machine Learning and the Physical
  Sciences Workshop at the 33rd Annual Conference on Neural Information
  Processing Systems}}.
\newblock 2019.
\newblock
  \href{http://www.arXiv.org/abs/2003.11603}{\texttt{arXiv:2003.11603}}.

\bibitem{Moreno:2019bmu}
E.~A. Moreno\hrefCMSnoop {}{ {et~al.}, ``{JEDI-net: a jet identification
  algorithm based on interaction networks}'',} \textit{ Eur. Phys. J. C}
  \textbf{ 80} (2020) 58,
  \href{http://dx.doi.org/10.1140/epjc/s10052-020-7608-4}{\doi{10.1140/epjc/s10052-020-7608-4}},
\href{http://www.arXiv.org/abs/1908.05318}{\texttt{arXiv:1908.05318}}.

\bibitem{Moreno:2019neq}
E.~A. Moreno\hrefCMSnoop {}{ {et~al.}, ``{Interaction networks for the
  identification of boosted $H \rightarrow b\overline{b}$ decays}'',} \textit{
  Phys. Rev. D} \textbf{ 102} (2020) 012010,
  \href{http://dx.doi.org/10.1103/PhysRevD.102.012010}{\doi{10.1103/PhysRevD.102.012010}},
  \href{http://www.arXiv.org/abs/1909.12285}{\texttt{arXiv:1909.12285}}.

\bibitem{Choma:2020cry}
\hrefCMSnoop {}{N.~Choma {et~al.}, ``{Track Seeding and Labelling with
  Embedded-space Graph Neural Networks}'',} in \textit{ 6th International
  Workshop Connecting the Dots 2020}.
\newblock 6, 2020.
\newblock
  \href{http://www.arXiv.org/abs/2007.00149}{\texttt{arXiv:2007.00149}}.

\bibitem{Bogatskiy:2020tje}
A.~Bogatskiy\hrefCMSnoop {}{ {et~al.}, ``{Lorentz Group Equivariant Neural
  Network for Particle Physics}'',}
  \href{http://www.arXiv.org/abs/2006.04780}{\texttt{arXiv:2006.04780}}.

\bibitem{Kieseler:2020wcq}
\hrefCMSnoop {}{J.~Kieseler, ``{Object condensation: one-stage grid-free
  multi-object reconstruction in physics detectors, graph and image data}'',}
  \textit{ Eur. Phys. J. C} \textbf{ 80} (2020) 886,
  \href{http://dx.doi.org/10.1140/epjc/s10052-020-08461-2}{\doi{10.1140/epjc/s10052-020-08461-2}},
  \href{http://www.arXiv.org/abs/2002.03605}{\texttt{arXiv:2002.03605}}.

\bibitem{Mikuni:2020wpr}
\hrefCMSnoop {}{V.~Mikuni and F.~Canelli, ``{ABCNet: An attention-based method
  for particle tagging}'',} \textit{ Eur. Phys. J. Plus} \textbf{ 135} (2020),
  no.~6, 463,
  \href{http://dx.doi.org/10.1140/epjp/s13360-020-00497-3}{\doi{10.1140/epjp/s13360-020-00497-3}},
  \href{http://www.arXiv.org/abs/2001.05311}{\texttt{arXiv:2001.05311}}.

\bibitem{Pata:2021oez}
J.~Pata\hrefCMSnoop {}{ {et~al.}, ``{MLPF}: Efficient machine-learned
  particle-flow reconstruction using graph neural networks'',}
  \href{http://www.arXiv.org/abs/2101.08578}{\texttt{arXiv:2101.08578}}.

\bibitem{Wang:2020fjr}
M.~Wang\hrefCMSnoop {}{ {et~al.}, ``{GPU-accelerated machine learning inference
  as a service for computing in neutrino experiments}'',} \textit{ Front. Big
  Data} \textbf{ 3} (2020) 48,
  \href{http://dx.doi.org/10.3389/fdata.2020.604083}{\doi{10.3389/fdata.2020.604083}},
  \href{http://www.arXiv.org/abs/2009.04509}{\texttt{arXiv:2009.04509}}.

\end{thebibliography}\endgroup
\end{document}